\documentclass[sigconf]{acmart}
\AtBeginDocument{%
  }

\copyrightyear{2026}
\acmYear{2026}
\setcopyright{cc}
\setcctype{by}
\acmConference[UIST '26]{The 39th Annual ACM Symposium on User Interface Software and Technology}{November 02--05, 2026}{Detroit, MI, USA}
\acmBooktitle{The 39th Annual ACM Symposium on User Interface Software and Technology (UIST '26), November 02--05, 2026, Detroit, MI, USA}
\acmDOI{10.1145/3830398.3830533}
\acmISBN{979-8-4007-2856-3/2026/11}

\newcommand{\xhdr}[1]{\vspace{1.7mm}\noindent{{\bf #1}}}

\usepackage{enumitem}
\usepackage{xspace}
\usepackage{setspace}
\usepackage{color,soul}
\usepackage{subcaption}
\usepackage[dvipsnames]{xcolor}
\usepackage[export]{adjustbox}
\usepackage[toc,page]{appendix}
\usepackage{listings}
\usepackage{tikz}
\usepackage{hyperref}

\lstdefinestyle{customlist}{
    basicstyle=\ttfamily\footnotesize,  
    numbers=left,                        
    numberstyle=\tiny\color{gray},       
    stepnumber=1,                         
    numbersep=5pt,                        
    showstringspaces=false,               
    breaklines=true,                      
    breakatwhitespace=false,              
    keepspaces=true,                      
    columns=fullflexible,                 
    frame=single,                          
    backgroundcolor=\color{gray!10}       
}

\newcommand{\hlc}[2][yellow]{{%
    \colorlet{foo}{#1}%
    \sethlcolor{foo}\hl{#2}}%
}
\newcommand\camera[1]{\textcolor{black}{#1}}
\newcommand\edit[1]{\textcolor{black}{#1}}

\newcommand{\stablelm}{\textit{stablelm-21-6b-chat}\xspace}
\newcommand{\gemmatwo}{\textit{gemma-2-2b-it}\xspace}
\newcommand{\mistral}{\textit{Mistral-7B-Instruct-v0.3}\xspace}
\newcommand{\qwen}{\textit{Qwen2.5-7B-Instruct}\xspace}
\newcommand{\gemmanine}{\textit{gemma-2-9b-it}\xspace}

\newcommand{\reducedstrut}{\vrule width 0pt height \ht\strutbox depth \dp\strutbox\relax}

\newcommand{\select}{%
  \begingroup
  \setlength{\fboxsep}{0pt}%
  \colorbox{SpringGreen}{\reducedstrut\texttt{SELECT}\/}%
  \endgroup
}

\newcommand{\learn}{%
  \begingroup
  \setlength{\fboxsep}{0pt}%
  \colorbox{Orchid}{\reducedstrut\texttt{LEARN}\/}%
  \endgroup
}

\newcommand{\calibrate}{%
  \begingroup
  \setlength{\fboxsep}{0pt}%
  \colorbox{SkyBlue}{\reducedstrut\texttt{CALIBRATE}\/}%
  \endgroup
}

\newcommand{\prompt}{%
  \begingroup
  \setlength{\fboxsep}{0pt}%
  \colorbox{lightgray}{\reducedstrut\texttt{PROMPT}\/}%
  \endgroup
}

\begin{document}


\title[Steerable Chatbots]{Steerable Chatbots: Exploring Personalization Control Interfaces via LLM Activation Steering}

\author{Jessica Y. Bo}
\authornote{This work was conducted as a Student Researcher at Google.}
\affiliation{%
  \institution{University of Toronto}
  \city{Toronto, Ontario}
  \country{Canada}
  }
\email{jbo@cs.toronto.edu}

\author{Tianyu Xu}
\authornote{Now at Google DeepMind.}
\affiliation{%
  \institution{Google AR}
  \city{Mountain View, California}
  \country{USA}
  }
\email{tyx@google.com}

\author{Ishan Chatterjee}
\affiliation{%
  \institution{Google AR}
  \city{Seattle, Washington}
  \country{USA}
  }
\email{ishanc@google.com}

\author{Katrina Passarella-Ward}
\affiliation{%
  \institution{Google AR}
  \city{San Francisco, California}
  \country{USA}
  }
\email{kpassarella@google.com}

\author{Achin Kulshrestha}
\authornote{Equal supervision.}
\affiliation{%
  \institution{Google AR}
  \city{Toronto, Ontario}
  \country{Canada}
  }
\email{kulac@google.com}

\author{D Shin}
\authornotemark[2]
\authornotemark[3]
\affiliation{%
  \institution{Google AR}
  \city{Mountain View, California}
  \country{USA}
  }
\email{deshin@google.com}

\renewcommand{\shortauthors}{Bo et al.}

\begin{abstract}
Personalizing LLM responses typically requires users to articulate their preferences through prompting, which can be burdensome at cold start and difficult to articulate in natural language. We introduce an alternative paradigm, steerable chatbots: rather than asking users to describe what they want, let them directly manipulate it via a linear factor. We implement this through activation steering, leveraging a linear scalar to control how strongly a preference is expressed in the LLM's output. We first assess the computational viability of \camera{activation} steering as a method to control granular preference expression, then we explore how the factor can be exposed to users. We prototype three \camera{activation} steering interface designs that vary on the axes of \textit{agency} (user-led vs. system-driven) and \textit{fluidity} (static vs. adaptive). A within-subjects user study (n=14) in cold-start personalization tasks shows the potential for steerable chatbots to align better with underlying user preferences than prompting alone, while revealing heterogeneous values around control, persistence, and transparency in LLM personalization. 
\end{abstract}



\received{20 February 2007}
\received[revised]{12 March 2009}
\received[accepted]{5 June 2009}

\begin{teaserfigure}
\centerline{\includegraphics[width=\linewidth]{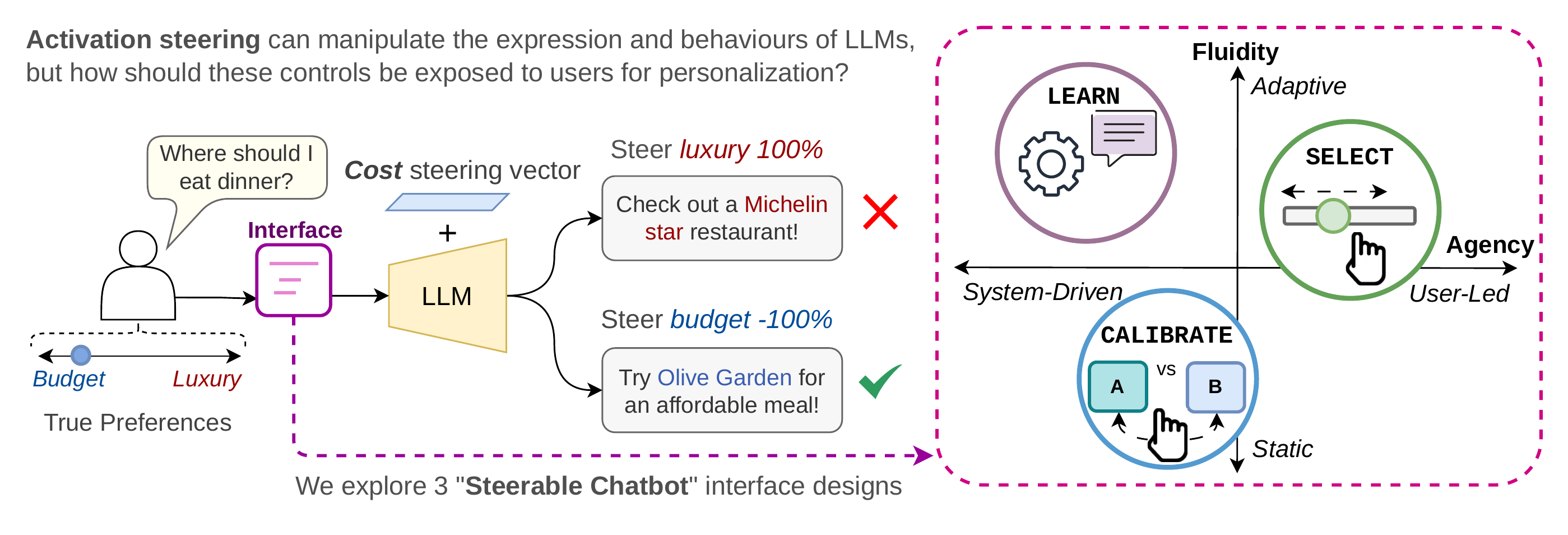}}
\caption{We explore how users can directly control the personalization of LLM-based chatbots to better align with their personal preferences, using a scalar factor to steer the model along a desired preference dimension. We \camera{evaluate} three designs of `steerable chatbots' that embody different personalization values of \textit{fluidity} and \textit{agency} -- \select{}, \calibrate{}, and \learn{}.}
\label{fig:teaser}
\end{teaserfigure}

\maketitle

\section{Introduction}
\label{sec:intro}
Large language models' (LLMs) ability to generate tailored responses to their users' preferences and needs lends them to be increasingly used as personal assistants \cite{li2024personal, xi2025rise}. In day-to-day planning tasks where preferences can be highly variable --- such as picking a restaurant for an anniversary dinner --- the LLM assistant needs to understand the user's underlying preferences and incorporate them in its response \cite{Chen2024-yg, kim2024understanding}. Currently, LLMs are trained to assume the preferences of an average user \cite{Jang2023-wa, DBLP:journals/tmlr/CasperDSGSRFKLF23}, while personalization relies on retrieving prior mentions of preferences from long-term memory \cite{lewis2020retrieval,DBLP:conf/ictir/SalemiZ25a, mysore2024pearl}, or well-specified prompts written by the user. The former is computationally intensive and does not support \textit{cold-start} personalization \cite{ryan2025synthesizeme}, while the latter demands structured prompting skills that many lay users lack \cite{zamfirescu2023johnny, tankelevitch2024metacognitive, oppenlaender2024prompting}. 

To shift this burden away from users, we explore the paradigm of \textit{directly steerable} chatbot interfaces for LLM personalization, provisioning users with manipulatable control over aligning the LLM with their latent preferences. In our steerable chatbots paradigm, users would not need to describe their preferences through prompting, as they can achieve their desired personalization through manipulating a linear steering factor. Figure \ref{fig:teaser} displays an example of how a question \textit{``Where should I eat dinner?"} can diverge significantly --- steering towards \textit{luxury} (\textbf{\textit{+100}}) results in high-end Michelin restaurants, while steering towards \textit{budget} (\textbf{\textit{-100}}) produces casual chain restaurants that are matched to the user. 



We leverage activation steering, a method for inducing desired behaviours in LLMs, and explore how it can be exposed through diverse user-facing controls for LLM personalization. Activation steering has emerged as a promising method for controlling LLM behaviours through injecting lightweight `steering vectors' into the internals of the model \cite{DBLP:conf/acl/SubramaniSP22, DBLP:conf/icml/LiuY0Z24, Turner2023-wy}. During forward inference, the pre-computed vectors are added to the representations at the residual stream of the LLM to modulate its outputs. The strength of \camera{activation} steering is set with a \textit{linear factor}, which controls the expression of the behaviour with a direction (positive or negative) and a magnitude. This factor is persistent, prompt-independent, and mappable to user-facing controls. \camera{Activation} steering provides computational benefits as it does not require any re-training of model weights like fine-tuning \cite{DBLP:conf/iclr/HuSWALWWC22}; it offers more granular control than natural language prompting \cite{DBLP:conf/acl/RimskyGSTHT24}; and it can be applied persistently or conditionally once the user's preferences are learned \cite{lee2025programming}. Although the technique has been primarily used in AI safety for model alignment, it creates an opportunity to expose the behavioural control of LLMs to users. \camera{Specifically, we adapt and apply the method developed by \citeauthor{DBLP:conf/icml/RutteABH24}}
 
Can \camera{activation} steering be used to help users improve their experience in tasks with high variability in preferences? 
We perform a exploration into user-controlled  \camera{direct manipulation interfaces} for LLM personalization, identifying opportunities where such steerable chatbots are most useful. 
In Section \ref{sec:framework}, we lay out how \camera{existing activation steering methods} can be used to parameterize personalization via controlling preference expressions. In Section \ref{sec:experiments}, we computationally assess its viability for aligning with varied preferences \camera{in day-to-day lifestyle planning tasks}. Finally, in Section \ref{sec:user}, we evaluate how users interact with three steering interfaces that embody distinct design values (\select{}, \calibrate{}, and \learn{}).
We contextualize our contributions within the domain of lifestyle planning, which is low-risk but has high variation in preferences across the population. 

\xhdr{Summary of Contributions.} 
\begin{enumerate}[topsep=0pt, itemsep=0pt]
    \item We introduce \textbf{steerable chatbots}, a personalization paradigm that exposes a \camera{direct manipulation LLM} control parameter to users via activation steering. We iterate within the \camera{interaction} design space of steerable chatbot interfaces and propose three diverse control interfaces.
    \item We validate the effectiveness of preference-based activation steering through computational experiments across five open-source LLMs \camera{in lifestyle planning tasks}.
    \item In an exploratory user study ($n=14$), we \camera{show} that steerable chatbots \camera{have potential to} improve preference alignment over an unscaffolded, prompting-only chatbot in cold-start \camera{personalization}.
    Qualitative findings discuss how individual values shape what users prefer in steering interfaces.
\end{enumerate}

\section{Related Works}
\label{sec:related}
The related works synthesize literature on personalizing LLMs, activation steering, and the design values for personalization controls.

\subsection{LLM Personalization}
Research on personalization for LLMs is vast, ranging from adopting personas to customizing content \cite{jiang2025survey, Chen2024-yg, Tan2023-eq, DBLP:conf/emnlp/Lu0QMCC23, DBLP:conf/acl/SalemiMBZ24, Allbert2024-gu, wang2024ai}. 
Technical methods for personalization have focused on storing unique user preferences in data-efficient manners, such as through retrieval-augmented generation \cite{DBLP:conf/ictir/SalemiZ25a}, model post-training \cite{Jang2023-wa, poddar2024personalizing, Li2024-fb, wang2024conditional, ryan2025synthesizeme}, and blending different LLMs or learning reward models \cite{pitis2024improving, chenpal, DBLP:conf/icml/0010PLQ00C24, chen2025pad, rame2023rewarded, Jang2023-wa, DBLP:conf/nips/ShiCHLHSD24, ryan2025synthesizeme}. These methods tend to require significant amounts of individual user data and would not be easily adaptable to cold start personalization scenarios. On the other hand, interface-driven methods focus on scaffolding the user interface to support better intent specification, such as through decomposing LLM outputs into sub-tasks \cite{ma2024beyond}, structuring the dimensions of the generation space \cite{DBLP:conf/chi/SuhCMLX24}, and interactively resolving ambiguities \cite{peng2025navigating, ma2025ambigchat}. These techniques tend to augment existing prompt-driven interactions, whereas we expose a direct manipulation linear steering factor for fine-grained control over preference expression. 

\subsection{Activation Steering}
Activation steering is a method for manipulating the internal states of LLMs at inference time to control the model's output without any model post-training. This is achieved by adding a steering vector --- computed by contrasting the desirable and undesirable representations of a behaviour --- to the residual stream of the LLM. Steering has been commonly used in AI safety to steer away from hallucinatory or harmful responses \cite{DBLP:conf/acl/RimskyGSTHT24, li2023inference, DBLP:conf/naacl/ZhaoDHDGWHWM25, DBLP:conf/iclr/BurnsYKS23, arditi2024refusal} and aligning with other desired model behaviours \cite{dong2023steerlm, DBLP:conf/iclr/StolfoBYHN25, DBLP:conf/icml/LiuY0Z24, cao2024personalized, kirk2025neural, karny2026neural}. The effect of steering can be further amplified, reduced, or even inverted by controlling the steering strength, a numerical factor that is multiplied to the vector. This property of \camera{activation} steering maps well to real-world preferences, which can be strong, weak, or in opposite directions. \camera{Currently, applications of activation steering for LLM personalization have not studied interactions with users, nor interface designs
\cite{konen2024style, he2025context, zhang2025personalized}}. 
\camera{Our work focuses on exploring the interaction design space of directly manipulatable LLMs, rather than improving the technical execution of activation steering. We adopt the existing steering method from \citeauthor{DBLP:conf/icml/RutteABH24} and apply it to a new downstream context of lifestyle planning, which is highly relevant to personalization.}

\subsection{Manipulatable Personalization Interfaces} 
Interfaces for human-LLM collaboration predominantly rely on text inputs \cite{kim2023cells, jiang2023graphologue, choi2024creativeconnect}, but the linear nature of the steering factor opens an opportunity for direct manipulation of model behavior. We organize the design space around two axes relevant to personalization \cite{segijn2021literature}: \textbf{agency} (how much control users have) and \textbf{fluidity} (how adaptively preferences are updated). The right-hand side of Figure \ref{fig:teaser} maps out how our interfaces fit within these dimensions, and Section \ref{sec:design_space} discusses the design rationale in more detail.

In the \textit{user-led and adaptive} design, we ground the \select{} interface, which provides users real-time control via a dynamic slider, with the principle of direct manipulation \cite{masson2024directgpt, shneiderman2010designing, chung2022talebrush}
This type of transparent and controllable personalization is well-supported: users value retaining explicit control over recommendations \cite{mcnee2003interfaces, knijnenburg2012inspectability, bakalov2013approach, louie2020novice, parra2015user} and generative outputs \cite{chung2023promptpaint, dang2022ganslider}.
As some users may find sliders to lack interpretability \cite{chung2023artinter}, the \textit{system-driven and persistent} design of \calibrate{} performs a calibration step to infer user preferences before the conversation begins.
This method draws on the framework of active learning \cite{rubens2015active}, Bayesian optimization \cite{yang2012bayesian}, and item response theory \cite{chen2005personalized}. 
Lastly, the \textit{highly-adaptive and system-driven} \learn{} interface operates without explicit user input, updating the steering factor in the background from conversational sentiment, analogous to feedback-driven approaches in LLM alignment \cite{DBLP:conf/iclr/ShaikhLHSCBY25} and recommender systems \cite{karabila2024bert, osman2019contextual}. This prioritizes an unobtrusive, temporally adaptive experience \cite{bauer2024values}.\

\section{Steerable Chatbots Paradigm}
\label{sec:framework}
In this section, we describe the proposed paradigm for steering-based LLM personalization, the design space for \camera{activation} steering control interfaces, and real-life applications of this personalization method. 
We target everyday LLM-assisted tasks with \textit{very high} variability in preferences, landing on \textit{lifestyle planning} as the task domain.

\subsection{Parameterizing Preference-Based Steering}
We describe the generalized personalization objective as follows. Given a task domain $T$ (e.g., \textit{lifestyle planning}) that necessitates highly variable solutions for each user, there is a set of interpretable preference dimensions, $\mathbf{d} = \{d_1, d_2, \dots\, d_n\}$. Each $d_i$ represents a preference dimension's strength, where a large positive value corresponds to a strong preference towards one end of the preference spectrum (e.g. \textit{luxury} for the dimension of \textbf{cost}) and a negative value to a preference towards the opposite end (e.g. \textit{budget}). The preference can also be uni-directional instead of bi-directional, through truncating the steering factor range with 0 as the minimum (e.g., \textit{neutral} $\rightarrow$ \textit{prefers seafood}).

An individual user $u$'s preference profile can be defined as $\mathbf{d^u} = \{d_{1}^u, d_{2}^u, \dots\, d_{n}^u\}$ and the preferred LLM response of user $u$ can be represented in a simplified manner as: $o = M(x, steer(\mathbf{d^u}))$. The generalized function $M(\cdot)$ represents the LLM's inference process, $steer(\cdot)$ represents the idealized \camera{activation} steering method that uses  $\mathbf{d^u}$ to modify the activations during inference, and $x$ represents the user's input to the LLM. The steering process described is $h_{steered} \leftarrow h + \mathbf{d^u} \cdot \mathbf{v}$, where $h$ represents the intermediate representations at the residual stream of the LLM, $\mathbf{v} = \{v_1, v_2, \dots v_n\}$ are the steering vectors associated with different preference dimension. Here, steering vectors capture desirable behaviours for the LLM to align with. By adding them into the residual stream during inference, alongside the \textit{strength} of steering represented by $\mathbf{d^u}$, the LLM's outputs are steered in the direction of the behaviour \cite{DBLP:conf/acl/RimskyGSTHT24}. 

While we present a model of multi-preference \camera{activation} steering here, we focus on steering one dimension at a time in our experiments. This decision to limit the complexity of the problem space decreases model instability from excessive steering, which would negatively impact the user studies. Under idealized conditions, however, multiple preferences reflecting a richer user profile could be steered simultaneously. Degeneracy and preliminary multi-dimensional results are discussed further in Appendix \ref{app:comp_results}.

\subsection{Design Space of Steering Interfaces}
\label{sec:design_space}
With preference-based steering defined, we explore the design space around how the linear factor can be exposed to users. We investigate how \camera{activation} steering enables different personalization interfaces across the axes of  \textbf{agency}, referencing how preferences are set; and \textbf{fluidity}, referencing when user profiles are learned and updated. Agency determines if the user has the ability to manipulate the parameters themselves, or if the personalization algorithm drives the learning in the background \cite{kaptein2015personalizing, jin2024implicit, maroto2024personalizing}. \camera{This narrows in on the tension between designs favouring higher user control \cite{bakalov2013approach, knijnenburg2012inspectability} versus those favouring higher automation \cite{jin2024implicit, maroto2024personalizing}.}
Fluidity determines if a user's preferences are  pre-\camera{computed} at the start of the interaction, or adaptively learned throughout the ongoing conversation \cite{chung2016adaptive, brusilovsky2007user}, \camera{which reflects how easily changing preferences can be accommodated.}

We \camera{developed three differentiated} interface designs across the two axes in order to broadly explore the design space. \camera{While there exist other axes of personalization, such as interpretability and granularity, we focus on these designs as preliminary exploration.}
These \camera{chosen} designs also \camera{resemble} existing interaction paradigms that users are generally familiar with, which reduces the learning curve of the experiment. The designs are visualized on the \textbf{agency}-\textbf{fluidity} axes in Figure \ref{fig:teaser}.
\begin{itemize}
    \item \select{} \textit{[High Agency, Medium Fluidity]}: Users directly select and update the steering factor at-will in conversation with the chatbot. The factor can also be statically set and treated as a part of the permanent user profile. This interface is similar to volume and temperature control sliders, which are adjusted by the user.  
    \item \calibrate{} \textit{[Medium Agency, Low Fluidity]}: Prior to entering the conversation, the user is prompted to iteratively select their preferred output produced by sampling different steering factors, and the factor is updated until their preference converges. This interface is similar to the gaze calibration step executed prior to using AR/VR headsets, where the individual parameters are pre-calibrated.
    \item \learn{} \textit{[Low Agency, High Fluidity]}: The steering parameter is updated in the backend of the chatbot based on feedback inferred in the user's response. To the user, the experience is similar to the typical interaction pattern with an AI chatbot. However, instead of personalizing by updating the memory bank with the user's written preference, the system makes an implicit inference based on the sentiment of the user's responses to update the steering factor.

\end{itemize}

\begin{table*}[t!]
\small
\caption{Bidirectional preference dimensions used in the experiments. The effect of the expressed preference is computed based on cosine similarity to a corpus of exemplary reviews from the Yelp Dataset, the processing details of which are provided. }
\label{tab:preferences}
\vspace{-0.75em}
\centering
\begin{tabular}{c|cc|c}
\toprule
\textbf{Preference} & \textbf{Negative} & \textbf{Positive} & \textbf{Yelp Dataset Processing Approach}   \\ 
\midrule
\textit{Cost}      & Budget            & Luxury            & Price point of 1 (budget) vs 4 (luxury).  \\ 
\textit{Ambiance}  & Touristy           & Hipster          & Ambiance of \textit{touristy} vs \textit{hipster}.   \\ 
\textit{Age}       & Kids            & Adults              & Attribute of\textit{ kids-friendly} being True vs False.      \\ 
\textit{Time}      & Evening           & Morning           & Business hours in the evenings vs morning.   \\ 
\textit{Cuisine}   & Asian          & American             & Keywords related to Asian vs American food. \\ 
\bottomrule
\end{tabular}
\end{table*}

\subsection{Application of Steerable Chatbots}
\label{sec:envisions}
\xhdr{Use Case.} Where would steerable chatbots be the most useful? We envision that they have the highest utility when users' preferences are persistent but difficult to articulate.
In cold-start scenarios, users cannot yet rely on conversational history to calibrate the LLM, and crafting prompts requires metacognitive effort that many users lack or find burdensome. 
\camera{Activation} steering addresses this by operating beneath the prompt layer: rather than asking users to specify what they want, it shifts the LLM's outputs along pre-defined preference dimensions. 
Steering can be applied continuously and uniformly across all outputs, functioning as a \textit{soft constraint} on the LLM's behavior that contextualizes every response without overriding the user's in-conversation input. 



\xhdr{Contextualizing Interactions with Prompting}. \camera{Activation} steering can be applied \textit{in tandem with} prompting. Steering and prompting serve fundamentally distinct goals in personalization --- the former aligns LLM outputs with users’ latent preferences, while the latter allows users to input additional requirements. Consider a low-income user who expresses, \textit{“I want to treat myself!”} in her request for restaurant recommendations. She may not realize that the LLM is interpreting her preference as luxury-focused and suggests a Michelin-starred restaurant out of her budget range. By applying a \textit{budget} steering vector, her prompted request would be better contextualized based on her latent cost preference, and her prompt simply shifts the chatbot's output to the higher end of her personal range.

\xhdr{Complex Preference Representation.}
We propose that user profiles can be represented through $\mathbf{d^u}$, a set of steering factors across pre-computed, interpretable preference dimensions. 
In the real world, personal preferences can be intricate and evolving, requiring more complex representations. For example, instead of grounding user profiles using the same dimensions, personalized steering vectors can be constructed from a user's own preference data. In long-term usage, the user can build up substantial interactions with the chatbot and provide positive and negative ratings for its responses (similar to \cite{ryan2025synthesizeme}). With high-quality data, it is possible to capture the nuances of an individual directly and dynamically update their personalized steering vector as new samples become available. 


\xhdr{Adapting to New Contexts.} Enabling steering in new task and preference domains is flexible. Given an open-source model where the internals are accessible, the steering vector of a desired concept can be computed from datasets that capture the desirable and undesirable LLM behaviours. These datasets can be sampled from real data, synthesized from LLMs, or even manually crafted by the user. Once computed, the vector is saved and cached, and then can be used with the user-defined strength factor during the inference of new outputs. The same steering dataset can be used for all downstream users, eliminating the need of user-specific data.

\section{Computation Experiments}
\label{sec:experiments}
Before studying interactions with users, we establish two technical preconditions for steerable chatbots: that steering reliably shifts expressed preferences (\textbf{E1},) and that it composes with user prompting (\textbf{E2}). We also conduct experiments exploring multi-preference steering (\textbf{E3} in Appendix \ref{app:e3}).
In this section, we describe task and preference dimensions, evaluation, technical implementation, and the results of the computational experiments. 

\subsection{Implementation and Evaluation of Steering}
To compute steering vectors, we follow \citeauthor{DBLP:conf/icml/RutteABH24}'s \edit{\cite{DBLP:conf/icml/RutteABH24}} method of training layer-wise linear probes that can accurately separate the representations of the positive and negative data samples. The coefficients of the logistic regression probe are used as the steering vector $v_i$, which can then be applied with the steering strength $d_i$ to amplify, dampen, or negate the preference expression in the LLM. More details about the technical implementation, including the parameter selection process, are provided in Appendix \ref{app:params}. For the dataset of positive and negative behavioural examples, we constructed the samples by prompting \texttt{GPT-4o} to output responses that would satisfy users of both the positive and negative sides of the preference. See Appendix \ref{app:steering_dataset} for examples.

We validate the steering procedure with five open-source LLMs available on Huggingface, ordered in increasing size: \stablelm (1.6B) \edit{\cite{bellagente2024stable}}, \gemmatwo (2B) \edit{\cite{team2024gemma}}, \mistral (7B) \edit{\cite{mistral7b}}, \qwen (7B) \edit{\cite{qwen}}, and \gemmanine (9B) \edit{\cite{team2024gemma}}. As the model's intermediate activations must be directly accessible, larger and closed-source models could not be used. All experiments are run on an NVIDIA L4 GPU. The hyperparameters are fixed with $temperature=0.7$, $top\_k=50$, and $top\_p=0.95$ to encourage some diversity through sampling. We constrain $max\_new\_tokens=100$ to restrict the length of the output. 
\begin{figure*}[th]
\begin{center}
\centerline{\includegraphics[width=\linewidth, left]{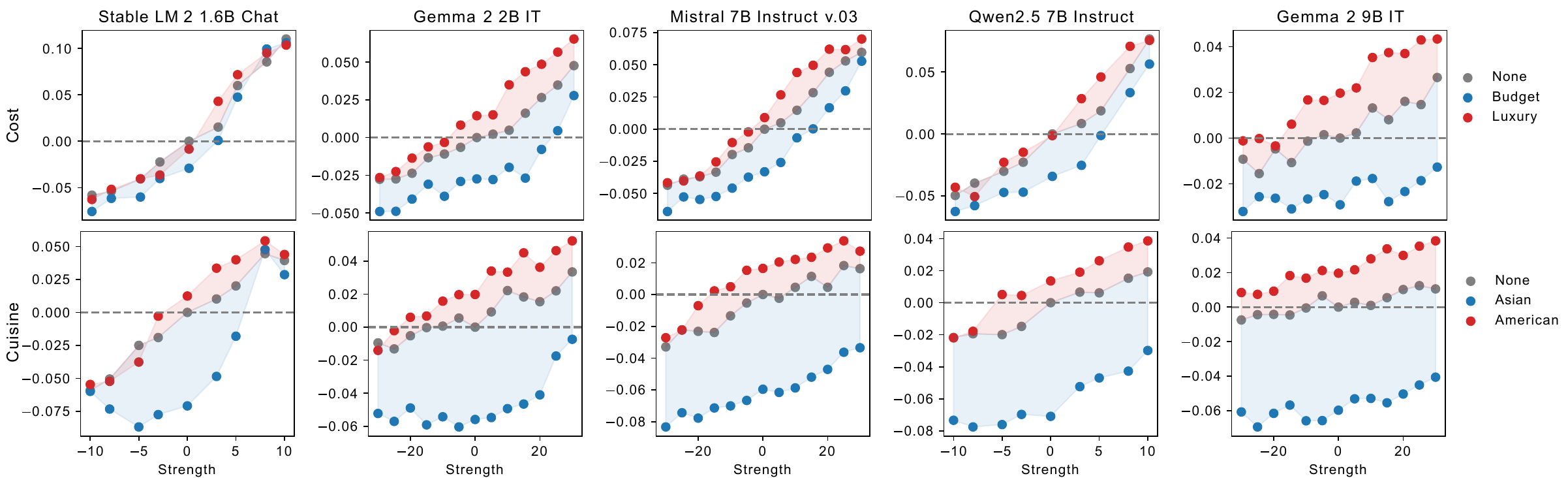}}
\caption{Interaction between prompting and steering for two preferences: \textbf{cost} (top) and \textbf{cuisine} (bottom) for \textbf{E2}.} 
\label{fig:prompting_effect}
\end{center}
\end{figure*}

\subsection{Evaluation of Preference Expression}
\label{sec:eval}
To evaluate steered LLM outputs against diverse, granular representations of human preferences in \textit{lifestyle planning} tasks, we use human-written Yelp reviews as a reference. We \camera{collect} reviews representing five common, interpretable preferences extractable from Yelp reviews metadata\footnote{\url{https://business.yelp.com/data/resources/open-dataset/}}: \textbf{cost}, \textbf{ambiance}, \textbf{age}, \textbf{time}, and \textbf{cuisine}, outlined in Table \ref{tab:preferences}. These dimensions are selected to show that \camera{activation} steering can generalize across preferences, but are not meant to be comprehensive. Each dimension is instantiated with bi-directional traits (e.g. \textit{luxury} and \textit{budget} for the dimension of \textbf{cost}), which correspond with the positive and negative directions of the steering vector. A limitation in this implementation is that nuanced and multi-dimensional preferences, such as \textbf{cuisine}, have to be minimized into bi-directional representations for simplicity. 

For both the computational and user experiments, we measure the \textit{effect of expressed preference }as the relative cosine similarity between the LLM output's embeddings and each of the bidirectional traits' reference Yelp reviews --- for example, the reference datasets for \textbf{cost} are sets of 700-900 positive reviews for establishments with a low and high price points, respectively. See Appendix \ref{app:yelp} for examples. $\textit{Effect} = \cos(e_o, \mathbf{\bar{e}}_+) -  \cos(e_o, \mathbf{\bar{e}}_-)$, where $e_o$ is the BERT embeddings of the LLM's output, ${\bar{e}}_+$ is the mean BERT embeddings of the positive trait's reference dataset, and ${\bar{e}}_-$ is that of the negative trait's reference dataset. All embeddings are computed using a pretrained Sentence-BERT model \cite{reimers-2019-sentence-bert} (\textit{sentence-transformers/stsb-roberta-base-v2} on HuggingFace.  \camera{We use cosine similarity as the measure following prior works that evaluate subjective constructs against a reference corpus \cite{an2018semaxis, kozlowski2019geometry}.} Additionally, similarity scores, rather than hard classification, allow the evaluation to capture more granular intensities of preference expression. 

We evaluate preference-based \camera{activation} steering on a dataset of $n=30$ real user queries sampled from the OASST2 dataset \cite{kopf2024openassistant}, which encompasses topics like travel planning, restaurant recommendations, recipe selection, and gift shopping.  We filtered questions that appeared at the start of an interaction and were open-ended and not grounded in particular preferences or specifications, like \textit{``What are the best restaurants in San Francisco?"} See examples of the queries used in Appendix \ref{app:queries}.

\subsection{Computational Experiment Results}

\subsubsection{(E1) Effect of \camera{Activation} Steering on Content and Quality. }
\textbf{How does steering strength affect the preferences expressed?} The objective of \textbf{E1} to establish a stable and functional steering range, where model outputs express the desired preference while maintaining response quality.
We apply steering for all models and preference dimensions with strength factors between $-30 < d < 30$, measuring the mean preference effect across the LLM's responses to the OASST2 query dataset. In addition to measuring the preference expression as described previously, we also compute the perplexity-normalized effect (PNE), which normalizes the expressed preference with the model's perplexity (a proxy for degeneration). If perplexity increases disproportionately more than the preference effect, then the PNE will be low, reflecting reduced quality. 

The correlation between steering strength and preference expression is shown in Figure \ref{fig:pnes} in Appendix \ref{app:e1}.  To summarize the findings, all models' preference expression is well correlated with the input steering strength, reflecting both the magnitude and direction of the steering factor.  However, the PNE is much lower outside the model's specific \textit{functional steering range} --- the approximate range of steering strengths for which steering does not significantly impact the LLM's output quality. 
We observe near-linear relationships between steering strength and perplexity-normalized preference effects
For all ensuing experiments, we only steer within the functional range of the LLM to preserve quality. The finding of \textbf{E1} is that \textit{\textbf{steering, in moderation, can effectively guide the preference expressed by LLMs}}.

\begin{figure*}[t!]
    \includegraphics[width=0.9\linewidth]{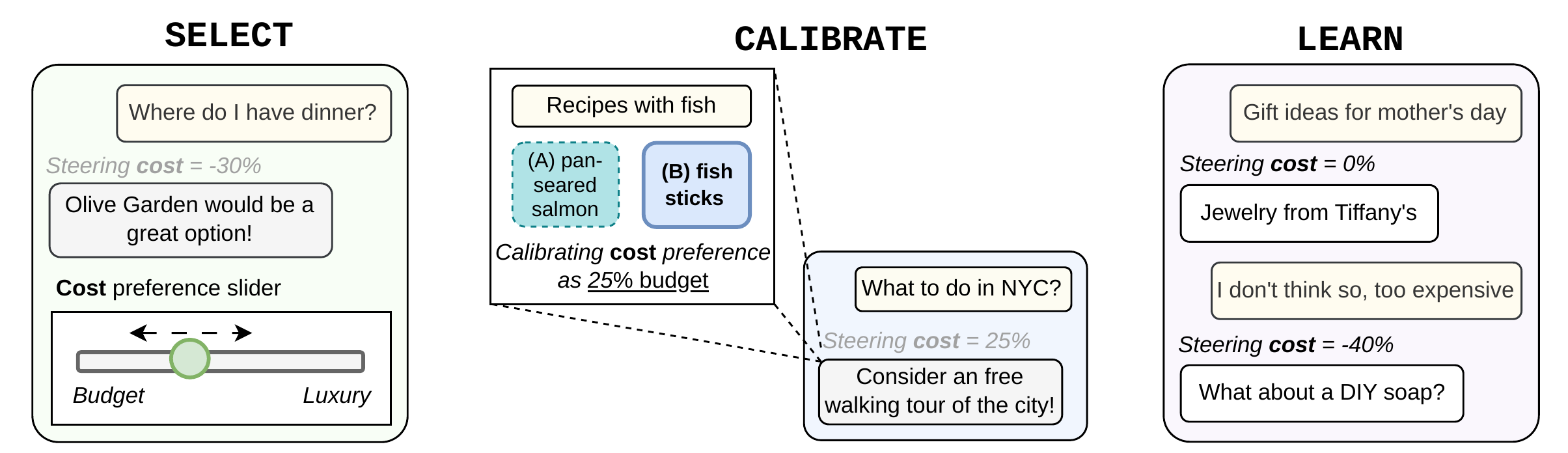}
    \caption{Graphical representations of the three steerable chatbot designs. See Figure \ref{fig:interfaces_screenshots} for screenshots of the real interfaces. }
    \label{fig:interfaces}
    \vskip -0.1in
\end{figure*}

\subsubsection{(E2) Effect of Prompting on \camera{Activation} Steering.}
\textbf{How does prompting interact with steering?} 
In our paradigm, steering is used to align with the user's latent preference profile, and the prompting adds contextualized variability --- e.g., \textit{“I want to treat myself!”} should yield different responses for luxury-seeking vs budget-oriented users when the LLM is correctly steered.
We repeat \textbf{E1}, but with the addition of preference-based prompts to each question in the query dataset. For each preference dimension, we augment the queries with prompting for both the positive and negative traits, such as \textit{``I am an early riser"} for the \textit{morning} trait and \textit{``I am a night owl"} for the \textit{night} trait. The objective is to understand how steering and prompting interact, especially what happens to the expressed preferences if the directions contradict or reinforce each other.

Figure \ref{fig:prompting_effect} depicts how adding preference-based prompts induces an \textit{offset} on the preferences expressed by the LLMs. Red represents prompting towards the positive trait, inducing a mostly positive offset; and blue represents the negative trait, inducing mostly a negative offset. We limit the main results to two dimensions (\textbf{cost} and \textbf{cuisine}) for brevity, but the other dimensions can be found in Figure \ref{fig:other_e2}, as well as specific examples in Table \ref{tab:prompting}, both in Appendix \ref{app:e2}.  Using the two-sample Kolmogorov-Smirnov test for significance ($p<0.05$), we find that for \textbf{cost}, only \gemmanine is significantly affected by prompting, while other models are not. For this reason, we use \gemmanine as the base model of the user study due to its sensitivity to prompting, which allows the model to respond to user prompts in addition to steering.

The significance of these results, while variable across models, indicates that \textbf{\textit{prompting can intensify or reduce the effect of preference expressions on top of \camera{activation} steering}}. This is a practical trait for a steerable chatbot --- steering to the user's underlying preference \textit{contextualizes} the overall conversation, and adding prompting introduces variability within the context.  

\section{User Study}
\label{sec:user}

With preference-based steering validated, we investigate how real users interact with and perceive steerable chatbots for personalization in an exploratory user study. We evaluate if \camera{activation} steering aligns chatbot outputs with users' ground-truth preferences (\textbf{U1}), and how users value the affordances of each steering interface design (\textbf{U2}). 
We compare the three different steering control interfaces with a prompting-only baseline with no added steering, \prompt{}.
The user study simulates cold-start conversations where the chatbot is not provided with any additional information about the user or scaffolding strategies specific to personalization. 

\begin{figure*}[t]
    \centering
    \begin{subfigure}[b]{0.47\textwidth}
    \centering
        \includegraphics[width=0.9\linewidth]{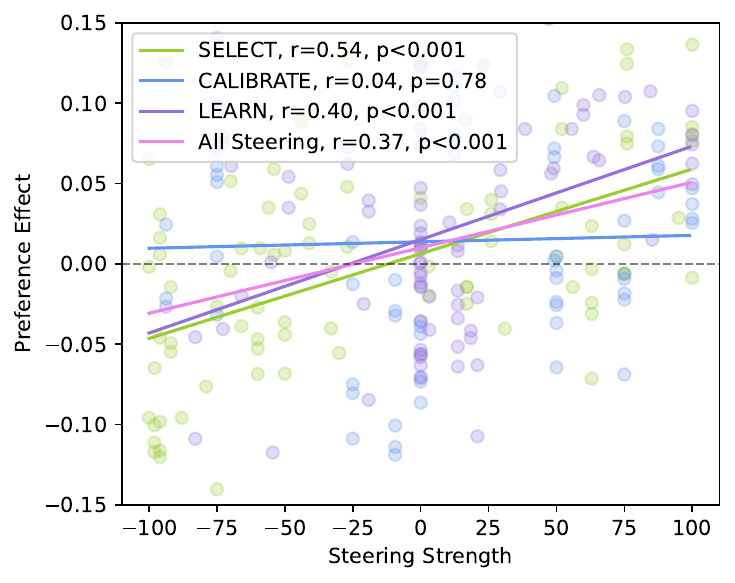}
        \caption{Correlation between the applied steering strength and the preference expressed in the chatbots' outputs.}
        \label{fig:steer_corr}
    \end{subfigure}
    \hfill
    \begin{subfigure}[b]{0.47\textwidth}
    \centering
        \includegraphics[width=0.9\linewidth]{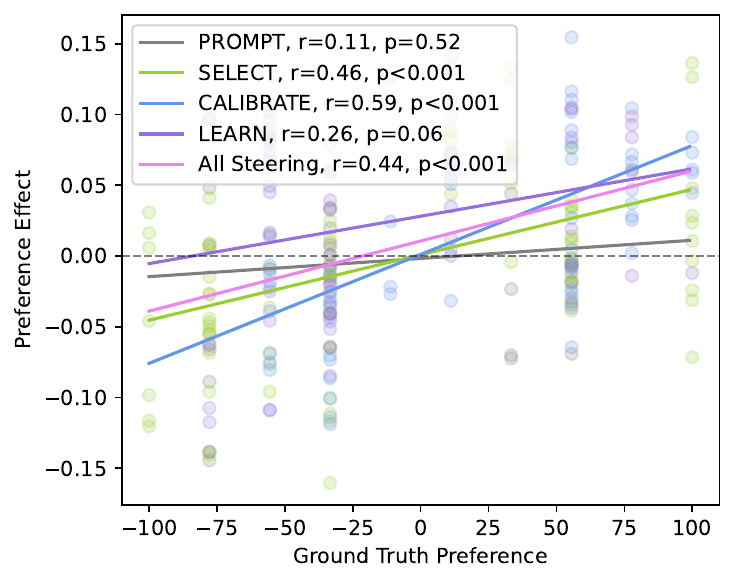}
        \caption{Correlation between the participant's ground truth preference and the preference expressed in the chatbots' output.}
        \label{fig:gt_corr}
    \end{subfigure}
    \vspace{-0.1em}
    \caption{Correlation analyses that demonstrate the steering produces preference-aligned outputs in the user study (\textbf{U1}).}
    \label{fig:correlations}
    \vspace{-0.1em}
\end{figure*}

\subsection{Implementation of Interfaces}
We implement all four chatbot interfaces using Gradio, an open-source Python package for prototyping visual interfaces of machine learning models \cite{abid2019gradio}. Participants were provided with a link hosting the chatbot interface, while the LLM (\gemmanine) is hosted in the backend. 
We standardize the preference strength range from -100 to 100 to make it interpretable for participants, but map it to \gemmanine's effective steering range. See Figure \ref{fig:interfaces} for a graphical representation of each interface, and Figure \ref{fig:interfaces_screenshots} for screenshots of the Gradio prototypes. 

\xhdr{\prompt{}}: The baseline interface has no added steering. Participants were encouraged to specify any personalization that they wish to add directly in the prompt. 

\xhdr{\select{}}: The steering parameter is controlled with a slider positioned at the bottom of the chat interface, with a range of -100 to 100. Participants are briefed that the selected slider value is instantaneously applied to the chatbot, and they can dynamically adjust it throughout the conversation.

\xhdr{\calibrate{}}: Prior to starting the conversation, users perform a calibration step to initialize the steering strength that is applied to the chatbot. Pairwise outputs (Response \textbf{A} and \textbf{B}) for are generated by sampling two different steering strengths.
Based on the user's preferred response, the steering strengths are updated to be closer to the preferred steering strength. For example, updating $d_A \leftarrow (d_A + d_B)/2$  if the preference is \textit{Slightly B}. This calibration step is repeated 2-3 times until the outputs \textbf{A} and \textbf{B} are roughly equivalent, with the final \textit{d} taken as the average of $d_A$ and $d_B$. 

\xhdr{\learn{}.} Similar to \prompt{}, the user can only control the personalization through prompting. To align the chatbot to the latent preferences of the user, we implement a simple learning algorithm that modifies the magnitude and direction of the steering factor based on the sentiment and preference direction of the user's latest message.
For example, \textit{"I don't like these gifts, they're too expensive"} would update the \textbf{cost} steering negatively towards \textit{budget}. The learned steering strength is displayed as a percentage (e.g. \textit{40\% budget}) in the interface for added transparency. 
For a message at timestep $t$, the update equation is given by Eq. (1), with implementation details in Appendix \ref{app:learned_strengths}.
\setlength{\abovedisplayskip}{3pt}
\setlength{\belowdisplayskip}{0pt}
\begin{equation}
\label{eq:update}
    d^{u*}_{t+1} \leftarrow d^{u*}_{t} + 
    p( \mathrm{dissatisfaction}(x_t) \cdot \mathrm{direction}(x_t) )
\end{equation}

\subsection{User Study Procedure}
We conduct a within-subjects study where participants used the prompting-only baseline and all three steering interfaces in randomized order to complete four lifestyle planning tasks. 
%
Each task corresponds to a preference dimension: gift shopping (\textbf{cost}), vacation planning (\textbf{ambiance}), restaurant recommendations (\textbf{cuisine}), and choosing meal prep recipes (\textbf{age}). Prior to starting the tasks, we collect participants' self-rated preferences on a scale of 1 to 10, where 1 mapped to the maximum negative preference and 10 mapped to the maximum positive preference. 
In each task, participants were instructed to engage in a conversation with the chatbot for at least several turns until they reached satisfactory responses, or until they exceeded 5 minutes in the task. 
Lastly, participants rated each chatbot on 7-pt Likert scale questions: 
\begin{itemize}[topsep=2pt]
    \item \textbf{Likelihood to Use}: \textit{How likely would you use the chatbot again for personalization tasks?}
    \item \textbf{Satisfaction}: \textit{How satisfied were you by the personalization?}
    \item \textbf{Perceived Control}: \textit{How in control did you feel of the personalization done by the chatbot?}
    \item \textbf{Perceived Persistency}: \textit{How well do you expect your preferences to be remembered by the chatbot for future use?}
\end{itemize}

The study concluded with a semi-structured interview where participants could openly discuss the interfaces that they liked and disliked, as well as what features and designs they considered to be important for personalized chatbots. To reduce biases, the order of conditions was randomized, the assignment of tasks to interfaces was counter-balanced, participants were not given the names of the chatbots, and participants were encouraged to honestly rate and discuss each interface.

\subsection{User Study Results}
\label{sec:user_results}
\subsubsection{Participants and Conversation Statistics}
We recruit 14 participants from the USA and Canada (\textit{women}=7, \textit{men}=4, \textit{non-binary}=1) with varied LLM use (\textit{daily}=7, \textit{weekly}=5, and \textit{monthly/occasionally}=2). The study lasted up to 60 minutes, for which participants were compensated with 25 USD gift vouchers. Figure \ref{fig:pref_dist} in Appendix \ref{app:gt_preferences} shows the participants' self-reported preferences for each of the four tasks and preference dimensions. For \textbf{ambiance} and \textbf{cost}, the preferences were well-distributed; for \textbf{cuisine} and \textbf{age}, the preferences were biased towards the negative and positive steering dimensions, respectively.

The number of turns of conversation for \prompt{} ranged from 2-10 ($M=5.25, SD=1.96$), \select{} from 4-10 ($M=6, SD=1.46$, \calibrate{} from 1-9 ($M=4.57, SD=1.76$), and \learn{} from 3-9 ($M=5.93, SD=1.98$). Using the Friedman test for differences between three or more non-parametric groups, we find that the lengths are not significantly different ($\chi^2(3)=5.75, p=0.12$). \calibrate{} conversations tend to be shorter because the participants spent more time at the calibration stage. The number of characters per message for \prompt{} ranged from 15-188 ($M=58.52, SD=27.40$), \select{} from 10-118 ($M=49.40, SD=23.68$), \calibrate{} from 9-119 ($M=54.34, SD=24.46$), and \learn{} from 15-188 ($M=58.41, SD=27.10$).  \select{} has shorter messages because participants did not specify their preferences in-prompt. Friedman test reveals a slightly significant difference ($\chi^2(3)=8.83, p=0.03$), but further post-hoc Wilcoxon signed-rank tests with a Holm-Bonferroni correction do not show significant differences between any pairs of conditions.

\begin{figure*}[t]
\vskip -0.1in
\begin{center}
\centerline{\includegraphics[width=0.75\linewidth]{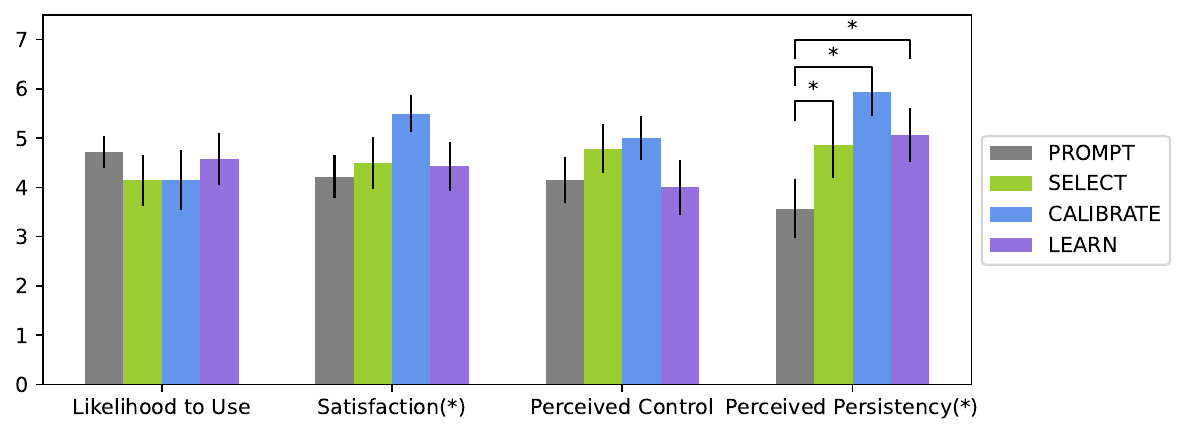}}
\caption{Post-experiment perceptions Likert ratings for all four interfaces. Significant contrasts between the steering conditions and \prompt{} are indicated by * in the figure. Perception categories that were rated significantly higher in participants' \textbf{favourite} interface are further indicated with (*) in the \textit{x-label}. }
\label{fig:perceptions}
\end{center}
\vskip -0.1in
\end{figure*}

\subsubsection{(U1) Aligning to Users' Latent Preferences}
Does \camera{activation} steering work in noisy, real user conversations? We examine if the method can a) \textbf{control the chatbot's expressed preferences}, and b) \textbf{produce preference expressions that align with users' latent preferences}. We measure the preference expression using the same method from Section \ref{sec:eval} and perform analyses with Pearson Correlation. The correlations between steering strength and the expressed preference for all steered chatbots are shown in Figure \ref{fig:steer_corr}. The correlation between participants' ground truth preferences and the chatbot's expressed preferences is shown in \ref{fig:gt_corr}. For this latter analysis, we dropped the first two turns of conversation for \prompt{} and \learn{} to allow them time to personalize to the user (as they both start off with no preset preferences). See Figure \ref{fig:learned_strenghts} in Appendix \ref{app:learned_strengths} for how the steering strengths were progressively learned over the conversations. 


We expect the correlation between steering \camera{strength} and preference expression in real conversations to be noisier than the computational experiments. Our results in Figure \ref{fig:steer_corr} show that \select{} ($r=0.54, p<0.001$) and \learn{} ($r=0.40, p<0.001$) both have \edit{moderate} positive relationships between the applied steering strength and the expressed preference, \camera{echoing the success of the computational findings}.
However, \calibrate{} was not correlated ($r=0.04, p=0.78$), \camera{likely due to the coarseness of the calibration step in capturing latent preferences accurately. To illustrate this failure:} \textbf{P2}'s calibration result \camera{mistakenly converged} at $cost=-75$, which reflects a \textit{budget}-oriented preference \camera{that they were ultimately dissatisfied with. To correct course, they} repeatedly prompted for more \textit{luxury} options \camera{in the conversation}  like, \textit \textit{"how about reputable or designer options"}. 
\camera{Combining together all variants of the} steerable chatbots, they reach a \camera{weakly} significant correlation ($r=0.37, p<0.001$). These results have two implications. One, \textit{\textbf{steering is effective at controlling the expressed preference}} when the user is satisfied with the personalization. Two, when the user is not satisfied, \textit{\textbf{prompting opposing preferences is effective at retaining control}} over the conversation \camera{(as it was for \calibrate{})}. 


\camera{Next, Figure \ref{fig:gt_corr} compares the chatbot's steered outputs and the users' ground truth preferences. In this analysis,} \select{} ($r=0.46, p<0.001$) and \calibrate{} ($r=0.59, p<0.001$) reach moderate, significant correlations, \camera{indicating that the chatbots were well-matched with their users' preferences}.
\learn{} \edit{($r=0.26, p=0.06$) fell short of a significant correlation, and \prompt{} was the worst ($r=0.11, p=0.52$)}. The sentiment-based update algorithm of \learn{} may not have been robust enough to \camera{capture diverse intentions from users' conversational feedback}. 
Overall, all \camera{variants of} steerable chatbots achieve moderate correlation with latent user preferences ($r=0.44, p<0.001$), outperforming \prompt{}. We further analyze the error between the steering strengths and the user's rated preferences in Figure \ref{fig:steering_stats} in Appendix \ref{app:error}, where the findings show that \select{} reaches the closest alignment, followed by \calibrate{}, then \learn{}.
While the correlation is moderate to small and the sample size ($n=14$) is limited, the findings cautiously suggest that \textit{\textbf{steerable chatbots  provide advantages over prompting-only chatbots in \camera{aligning with user preferences in} cold-start personalization}}. We recommend validating these results in larger, more ecologically valid studies.

\subsubsection{(U2) Subjective Perceptions of Interfaces}
As each interface embeds different design values, we triangulated between participant's Likert responses with their richer, descriptive interview data to understand the affordances provided by each steerable chatbot. 
The post-experiment Likert ratings are plotted in Figure \ref{fig:perceptions}. For every perception category, we test planned contrasts between the baseline \prompt{} with each steerable chatbot condition, using the one-sided Wilcoxon signed-rank test. We apply a Holm-Bonferroni correction to the \textit{p-value} to account for the multi-hypothesis testing. Strikingly, the first immediate observation is that there is no significant difference between interfaces in the categories of \textit{likelihood to use}, \textit{satisfaction}, and \textit{perceived control}. For \textit{perceived persistency} of the personalization, \select{} ($W=37.5, p=0.04$), \calibrate{} ($W=55.0, p=0.002$), and \learn{} ($W=36.0, p=0.006$) were all rated significantly higher than the baseline, \camera{indicating that steerable chatbots are perceived to retain their personalized settings across future sessions}. 

Examining the ratings more closely, we observe that participants often gave strong ratings towards \textit{one} steering interface, but not the others. As an added analysis, we operationalize each participant's \textbf{favourite} interface
as the one which received the highest rating in the \textit{likelihood to use} category. This results in a count of 6 votes for \select{}, 5 for \learn{}, and 3 for \calibrate{} --- where the highest rated interface is a steered chatbot for all participants, never the baseline. With a one-sided paired sign test, we find that the \textbf{favourite} interface is rated significantly higher in \textit{satisfaction} ($S=0.77, p=0.04$) and \textit{perceived persistency} ($S=1.0, p=0.001$), but just short of significance for \textit{likelihood to use} ($S=0.75, p=0.07$) and \textit{perceived control} ($S=0.8, p=0.05$). 
The discrepant ratings across the steerable chatbots suggest that what users want from personalization is not uniform. 

\begin{figure}[t]
\vskip -0.1in
\begin{center}

  \begin{subfigure}[b]{\linewidth}  
    \centering
    \includegraphics[width=0.75\linewidth]{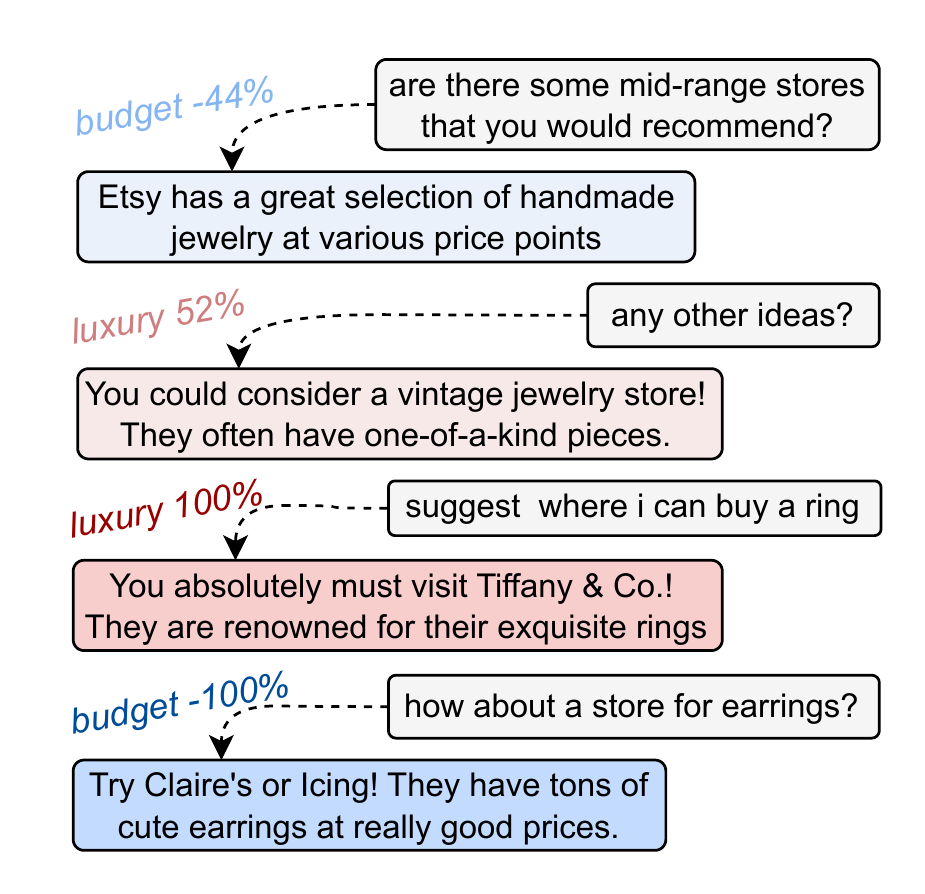}
    \caption{\camera{P7's} conversation for the \textit{Gift Shopping} task, where they changed the steering strength for \textbf{cost} using the \select{} interface.}
    \label{fig:select_convo}
  \end{subfigure}
  \begin{subfigure}[b]{\linewidth}  
    \centering
    \includegraphics[width=0.8\linewidth]{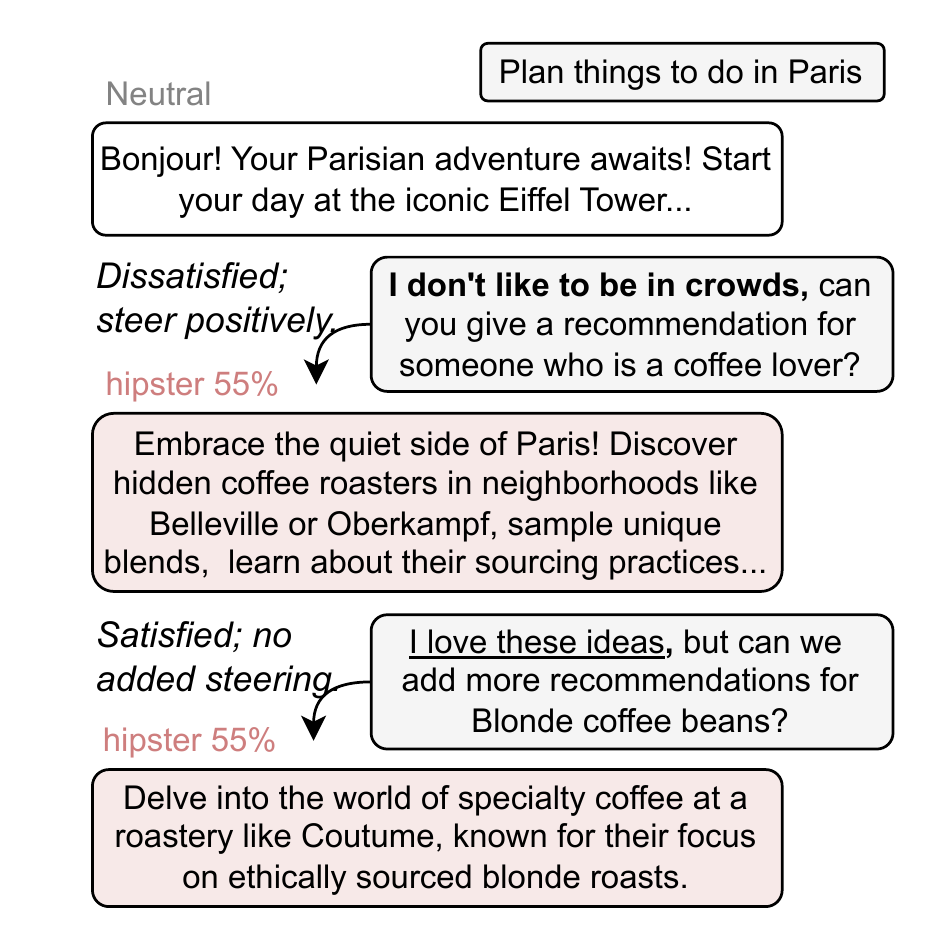}
    \caption{\camera{P10's} conversation for the \textit{Travel Planning}, where \learn{} updated the \textbf{ambiance} steering strength based on the sentiment of the user's messages (\emph{bold} for negative, \underline{underline} for positive). }
    \label{fig:learn_convo}
  \end{subfigure}
  \caption{Two sample user conversations for \select{} and \learn{}}
\label{fig:sample_convos}
\end{center}
\vskip -0.1in
\end{figure}

For nuanced understanding, we turn to the interview data, which reveals that participants' reactions to each interface were shaped by underlying values that map directly onto the axes of our design space --- \textbf{how much \textit{agency} they wanted over the personalization, and how \textit{fluidly} they expected it to adapt}.
We thematically analyze the post-task interview using an inductive coding approach \cite{clarke2017thematic}, shaped by themes related to the design values of the interfaces. Two researchers coded quotes extracted from the interview transcripts and refined the codes into agreed upon themes. We uncover marked heterogeneity in people’s preferences about personalization and disseminate them through the following findings.

\xhdr{Theme 1: \textit{\camera{Controlled} vs. Deferential Personalization.}} 
We first identify contrasting values in how much control users desire in personalization.
Participants who value control prefer \select{}, which allowed them to directly manipulate the chatbot. \textit{“I was more in control of the kind of answers I wanted to see”} said P2, and P4 agreed with \textit{“whether you felt like you were in control was extremely important”.} 
Users liked being able to use the slider to nudge the alignment of the chatbot's preference expression while communicating more granular requirements through prompting. P4 observes, \textit{“I can control this dimension [via the slider] and then I can tune the other parts that I want via conversation”}, and P6 echoes that it was like \textit{“‘drawing a box’ around the personalization factors that I want to explore”}. \camera{Users could also be fully alleviated from the mental load of prompting, like P7 who exploited the slider on \select{} to nudge the chatbot's response towards \textit{luxury} without having to specify their preferences in-prompt (see the conversation in Figure \ref{fig:select_convo}). }


On the other end of the spectrum, some users \camera{like deferring} to the system, relieving them of making the personalization choices themselves.
P12 believed \learn{} was more straightforward and expressive, \camera{with wanting to} \textit{ “let me just dump my thoughts and let the LLM handle what's going on”}, and P14 agrees with saying they prefer to \camera{let the chatbot} \textit{“adjust my preference through multi-turn prompting”}. In terms of the user experience, P8 said \textit{“learning from conversations feels a little bit more flowy and natural”}, and P9 echoes with \textit{“it was more seamless and I felt like I could best express what I was looking for in more detail”}. \camera{P10 expressed being pleasantly surprised by how well \learn{}'s steering strength for \textit{hipster} adapted to the their preference to \textit{not be in crowds} (see Figure \ref{fig:learn_convo} for the conversation).}
However, participants who did not like \learn{} pointed out the lack of control over what the system learns. P4 expressed feeling judged, especially \textit{“if [the chatbot] thinks that I have a certain preference but that's not what I intended”}. \camera{P11 wants the learned steering factor to be adjustable and P13 agrees that \textit{“what a chatbot knows about me”} should be editable.}


This summarizes the trade-off on the \textbf{agency} axis, where many prefer to retain control with the \textit{user-led} \select{}, but others acknowledge that the \textit{system-led} design of \learn{} offers seamless personalization via conversation. In either case, users indicate that they want to understand the personalization applied and be able to modify the information when necessary, \camera{highlighting the importance of transparency.}

\xhdr{Theme 2: \textit{Situational Personalization Persistence.}} 
Overall, participants generally found value in personalization that persists over sessions. This is reflected in the significance of ratings in the \textit{perceived persistency} category in Figure \ref{fig:perceptions}. In the interview, P3 said the baseline \prompt{} interface felt like having \textit{“a random conversation with a stranger”}, expressing dissatisfaction in having to describe their preferences. However, people held different views on how stable personalization should be. Some, like P9, valued the ability to update their preference through \select{}, while others, like P3, believed their preference \textit{“is probably set for the entire conversation”} and does not need to be changed. \camera{In general, we find disparate orientations towards the permanence of preferences.}  P14 believes that \textit{“people's thoughts can change all of a sudden”}, even throughout the same conversation. P6 acknowledges that \textit{“preferences tend to change over time”}, \camera{hesitantly saying that \textit{“I don't know if I actually want my chatbots to remember my preferences for future use”}}. P13 notes that they only want \textit{“personalization for certain things, but not others}” \camera{--- referring to a separation between their work and personal profiles, where the preferences would diverge. These examples showcase how a seemingly desirable trait, like persistent personalization, should be be approached with contextual nuance.}

Our personalization interfaces support different levels of \textbf{fluidity}. \learn{} and \select{} allows lightweight and dynamic preference updates, while \calibrate{} preferences are fixed after the calibration step. 
Users who liked \calibrate{} \camera{generally believed that} their preferences are more static, and the calibration step \camera{has the advantages of saving time and effort}. P3 \camera{reports enjoying the calibration steps, saying} \textit{“\camera{I'll be less frustrated while [using the chatbot], because }I've already personalized it up to some extent with my liking”}. P8 also liked \textit{“having the [personalization] set beforehand…that way I don't have to think about it and have a back-and-forth conversation”}.
The focused, interactive calibration questions also invoked a perception of precision, with P10 saying \textit{"it kind of started to narrow down... so I feel like that was understanding very detailed preferences"} and P7 comparing it to an eye doctor appointment, \textit{"is it option A or option B?.. eventually they get to your prescription"}. This is \camera{empirically} reflected in the correlation analysis in Figure \ref{fig:gt_corr}, showing that \calibrate{} matched \camera{ground truth} user preferences the best. 


We find that for \textbf{fluidity}, \camera{higher} \textit{adaptivity} is particularly valued if the inference for user preferences is erroneous, such as in the \textit{system-driven} interfaces. P9 asserts that \textit{“if [a chatbot] was making a false inference that was going to lead to more inaccurate outputs, I would want to be able to stop that in its tracks”}. This sentiment highlights what many felt --- if the personalization is not helpful or accurate, there should be a way to change it. However, if preferences can be accurately inferred with overhead steps, some users appreciate persistence in personalization. 


\xhdr{Theme 3: \textit{Interpretability of steering.}}
\camera{Outside of the design space dimensions}, participants also brought up general feedback on the interpretability of preference-based steering. 
Participants were split over the transparency feature of \learn{}, where the learned steering strength is exposed in the conversation. P12 found the transparency helpful, \textit{“I liked being able to see how the model was adjusting”}, and P14 said it helped them understand the \textit{“chatbot’s mental model of the user's preferences}”. However, P11 said the preference percentage \textit{“was overloading me...honestly I was ignoring it”}, \camera{and P4 said \textit{“it felt judgy”}.} For \select{}, \camera{P5 found the numerical steering strength of the slider to lack interpretability, critiquing that} \textit{“it gets into the nuance of how [the chatbot] understands those [preferences] and how you understand them”}. P11 believed that \textit{“people might not know how to \camera{[use the slider]} correctly for more complex dimensions”}, \camera{reflecting how interpretability challenges will scale up}. The strength of \calibrate{} is clear here, with P5 saying, \textit{“instead of just picking a word that you and the AI are defining differently, it can ask you questions”}. This shows that the interpretability of preference strength should be improved, such as by providing more in-context examples and calibrating users to the correct mental model ahead of the conversation.

\section{Discussion}

\subsection{Key Findings}
In this work, we propose a paradigm for LLM personalization using activation steering, where relevant preference dimensions for a task can be amplified or dampened with pre-computed steering vectors moderated by linear strength factors. Our computational experiments validate the generalizability of steering across five different models and preference dimensions \textbf{(E1)}, and its effectiveness in tandem with prompting \textbf{(E2)}.
Our exploratory user study further suggests that steering-based chatbots can align better with latent user preferences than a prompting-only chatbot in cold-start personalization \textbf{(U1)}. We explore how the steering factor can be exposed to users via three interface designs, guided by a design space of agency and fluidity. Participants report highly divergent preferences shaped by personal values, highlighting trade-offs among control, persistence, and transparency --- for which our steerable chatbots design space offers structured exploration \textbf{(U2)}.

\subsection{Implications and Future Work}
\xhdr{Interfaces for Interacting with Chatbots.}
The key difference between steerable chatbots and prompting-based chatbots lies the interaction affordances enabled through the linear steering factor. 
Methods for interacting with chatbots outside of natural language prompting are rising in popularity, including graphical interfaces that can promote structured exploration across diverse spaces \cite{DBLP:conf/chi/SuhCMLX24}, improve prompt structuring \cite{ma2024beyond}, manipulate and refine ideas \cite{masson2025textoshop, masson2024directgpt}, and sketch relationships between components \cite{yen2025code, rosenberg2024drawtalking}. 
Enabling different modalities of inputs, control, and sensemaking can be highly transformative in supporting complex LLM workflows. 

\xhdr{Personalizing the Personalization Experience.}
The variance in participant feedback on the steerable chatbot interfaces indicates highly heterogeneous preferences towards how personalization is implemented. \camera{We found that CALIBRATE suits users with stable preferences; LEARN is liked by users whose preferences shift mid-conversation; and SELECT fits users who want explicit, ongoing control. }Among the main themes that participants expressed in the interviews, values related to \textit{agency} and \textit{fluidity} where echoed in explanations about which chatbots participants liked most. In particular, the need to understand how personalization is performed and retain control over what is learned and stored stands out, and this is reflected in the findings of prior research  \cite{mcnee2003interfaces, knijnenburg2012inspectability, bakalov2013approach}. Ultimately, meta-preferences within personalization may be determined through a combination of individual and task factors. Blanket designs (including the unscaffolded, prompt-only chatbots) are not always suitable --- pointing towards malleable interfaces that can be customized to suit personalized experiences \cite{cao2025generative}. 

\xhdr{Viability of \camera{Activation} Steering as LLM Control.}
Activation steering, as a technical method of LLM control, offers both benefits and limitations.
A strength of steering is that it requires only the inference-time application of a pre-computed vector; it is incredibly resource-efficient in comparison to methods that require updating model weights or storing extensive user data \cite{DBLP:conf/acl/RimskyGSTHT24}. Since the steering vector and strengths contain no personally-identifiable data (unlike personalized finetuning weights or explicit user profiles), it is also more sensitive to user privacy concerns \cite{aguirre2016personalization, ischen2019privacy}. However, the question of preserving generation quality and robustness is important to address \cite{DBLP:conf/icml/WuAG00JMP25, DBLP:conf/icml/RutteABH24}, and more stability is needed of the method to adapt it to steering complex personalization setups. 

\xhdr{Generalization to Long-Term Personalization.}
In our focus on cold-start personalization, we chose to parameterize steering via a top-down approach, relying on pre-defined preference dimensions where every user can be modelled as a simple vector $\mathbf{d^u}$, rather than requiring prior interaction data. 
In the long-term, steering can be enhanced via integration with other personalization strategies like storing user details in RAG systems \cite{DBLP:conf/ictir/SalemiZ25a}. Alternatively, steering can also be used in a bottom-up, data-driven approach, where the user's own data (such as likes and dislikes) is used to create the personalization signal. This is viable given comprehensive user preference data collected from diverse scenarios over prolonged interactions. \camera{This approach is more generalizable to complex preferences, but would require substantial individual data.}


\subsection{Limitations}
We conduct the user study with low-fidelity prototypes, where the calibration and learning algorithms were meant to convey proof-of-concept capabilities. 
For \learn{}, this limited the accuracy of the learned preferences, which impacted user satisfaction in the steered outputs. 
\camera{For \calibrate{}, the ambiguity of the underlying preference sometimes resulted in misalignment with users' interpretations. Complex preferences had to be reduced down to singular dimensions (although we do perform preliminary multi-dimensional steering in Appendix \ref{app:e3}), limiting the ecological validity of how preferences are represented.} 

Steering was implemented using a simple method \cite{DBLP:conf/icml/RutteABH24} with small, local models, which do not reflect the personalization potential of state-of-the-art chatbots.  As such, we asked the participants to concentrate on the interaction modality of the interfaces in their post-task interviews, rather than the personalization quality. 
The prompting-only baseline was implemented without sophisticated prompting interventions, such as providing structured templates for expressing preferences, as it was meant to approximate a naive, unscaffolded chatbot. Accordingly, our results showing that steering improves personalization to latent user preferences should therefore be taken in the context of a relatively weak baseline.

Our low-stakes task domain does not represent all personalization use cases, particularly more critical domains like \textit{value alignment} \cite{DBLP:conf/icml/SorensenMFGMRYJ24}. Future explorations can include tasks such as creative writing \cite{gomez2023confederacy}, communications \cite{yang2024talk2care}, and persona adoption \cite{Ha2024-iq, Chen2024-yg}. We were also constrained to a set of five preferences based on their relevance to the task and their presence within the Yelp Dataset, \camera{which represents a relatively narrow preference subspace as the dataset is dominated by restaurant reviews. Our evaluation setup therefore may not apply to tasks outside of the lifestyle planning domain.}
With the advancement of \textit{LLM-as-Judge} evaluation capabilities \cite{gu2026survey}, future evaluation techniques may not be constrained by existing human preference datasets \cite{chen2025persona}.

\section{Conclusion}
\label{sec:conclusion}
We demonstrate how activation steering can be applied towards personalizing LLMs in preference-driven tasks, with both computational and user studies. We develop three interaction interfaces for \textit{steerable chatbots} based on design values relevant to personalization, and compare their results in aligning chatbot outputs with latent user preferences and subjective perceptions from end users. 
Our findings reframe LLM personalization as an interaction design problem and showcase activation steering as a resource-efficient mechanism for driving directly manipulatable chatbots.

\clearpage


\bibliographystyle{ACM-Reference-Format}

\bibliography{references}

\newpage

\appendix
\onecolumn

\renewcommand\thefigure{\thesection.\arabic{figure}} 
\setcounter{figure}{0}  

\renewcommand\thetable{\thesection.\arabic{table}} 
\setcounter{table}{0}  

\renewcommand\thelstlisting{\thesection.\arabic{lstlisting}} 
\setcounter{lstlisting}{0}  

\setcounter{secnumdepth}{3}


\section{Implementation of Preference-Based Steering}
\label{app:implementation}
This appendix describes the technical steering parameters and datasets that were used in the computational experiments in Section \ref{sec:experiments}. 

\subsection{Steering Parameters}
\label{app:params}
We followed \citeauthor{DBLP:conf/icml/RutteABH24}'s procedure for selecting the probe type (in additional to the logistic regressor, we also tested difference-in-means and principal component analysis) and the \textit{k} hyperparameter for the \textit{top-k} layers, where steering is selectively applied to the \textit{top-k} of the layers with the highest detection accuracy.
Table \ref{tab:params} shows the best performing parameters that were used for each model. \
 We also experimentally determined an approximate functional steering range where the model is not affected significantly by degeneracy. 

\begin{table}[h]
\centering
\caption{Steering parameters selected for each model.}
\vspace{-0.5em}
\label{tab:params}
\begin{tabular}{cccc}
\toprule 
\textbf{Model}                     & \textbf{Top K Layers} & \textbf{Functional Steering Range} & \textbf{Probe} \\ \midrule
\textit{\stablelm}  & 16                    & (-10, 10)                 & Logistic       \\
\textit{\gemmatwo}  & 16                    & (-30, 30)                 & Logistic       \\
\textit{\mistral}   & 24                    & (-30, 30)                 & Logistic       \\
\textit{\qwen}      & 24                    & (-10, 10)                 & Logistic       \\
\textit{\gemmanine} & 32                    & (-30, 30)                 & Logistic      \\
\bottomrule
\end{tabular}
\end{table}

\subsection{Steering Dataset}
\label{app:steering_dataset}
Due to a lack of LLM output datasets on preferences, all steering datasets are generated by \texttt{GPT-4o} to emulate what the gold standard of embodying this preference should look like -- e.g \textit{``Generate a list of responses that a chatbot might reply to a user who is looking for \textbf{kid-friendly} suggestions across user queries in lifestyle planning tasks such as travel planning, restaurant recommendations, recipe selections..." } See Listing \ref{lst:steering} for sample instances from the positive and negative steering dataset for \textbf{age}. Across all steering datasets in the five preference dimensions, we generated 50-80 samples per trait. We explored alternative datasets such as a list of synonymous adjectives (e.g.,\textit{cheap} suggestions, \textit{affordable}, \textit{low-cost}... for \textbf{cost}) and the Yelp reviews, but the former did not have an appropriate format to guide LLM outputs and the latter is too noisy to learn meaningful concepts from.
\begin{lstlisting}[style=customlist, caption={Sample instances from the steering dataset for \textbf{age}}, label={lst:steering}]
[Positive steering dataset for adults-oriented]
Visit the rooftop bar downtown for stunning views and a curated selection of craft cocktails.
Plan a day of wine tasting at the region`s top vineyards, complete with private tours.
Book a spa day with massages, saunas, and relaxation areas exclusively for adults.
Attend a live jazz performance at the speakeasy-style venue known for its intimate ambiance.
Join a mixology workshop to learn the secrets behind making perfect cocktails at home.
Take a sunset cruise offering drinks and live music in a serene, adults-only setting.
Visit the cigar lounge downtown, featuring an extensive selection and a cozy atmosphere.
Book a brewery tour with tastings of seasonal beers and behind-the-scenes insights from brewmasters.

[Negative steering dataset for kids-oriented]
Visit the interactive science museum with hands-on exhibits perfect for kids of all ages.
Plan a day at the local zoo, featuring kid-friendly animal encounters and feeding sessions.
Explore the city`s largest playground with climbing structures, slides, and picnic areas for families.
Spend time at the local aquarium, known for its touch tanks and playful sea otters.
Visit the city`s amusement park with rides designed specifically for younger kids and toddlers.
Head to the local library for storytime sessions and engaging activities for kids.
Visit the butterfly garden - it`s colorful, educational, and a hit with younger children.
Take a scenic train ride that offers kid-friendly entertainment and stunning views along the way.
Visit the local zookeeper for a behind-the-scenes tour - kids love learning about animals up close.
\end{lstlisting}

\subsection{Evaluation Queries Dataset}
\label{app:queries}

Listing \ref{lst:queries} shows a sample of real user queries selected from the OASST2 dataset in the domain of \textit{lifestyle planning} that were used throughout the computational experiments. These queries were manually identified and filtered by the research team. 
\begin{lstlisting}[style=customlist, caption={Sample task queries used in the computational experiments.}, label={lst:queries}]
Give a recipe idea for a vegetarian meal which has tofu?
what are some popular souvenirs people take home when they visit the usa from europe?
What are the best restaurants in San Francisco?
Hi, I am trying to plan a camping holiday with my family. What supplies do you suggest I purchase or pack?
I want to buy a gift for my gir friend for the valentines day
What are some things I should do on a 5-day vacation in Thailand?
What is the best way to cook a tomato?
...
\end{lstlisting}


\subsection{Reference Yelp Reviews for Preference Effect Computation}
\label{app:yelp}

Listing \ref{lst:hipster} shows samples of the reference Yelp reviews datasets for \textit{hipster} and \textit{touristy} of the \textbf{ambiance} preference dimension. The samples are first collected based on the criteria listed in Table \ref{tab:preferences}, then filtered down for relevancy to the trait through a combination of manual human evaluation and automated LLM evaluation with \texttt{GPT-4o-mini} as the judge.

\begin{lstlisting}[style=customlist, caption={Sample Yelp reviews of the traits \textit{touristy} and \textit{hipster} for the \textbf{ambiance} dimension (some are abbreviated for length).}, label={lst:hipster}]
[Yelp reviews for hipster]
There are restaurants that you go to because they are unique enough to stand out from anything in their category
I would highly recommend if you are looking for somewhere different to eat that does not have the usual bar fare. 
I wish I wasn't reviewing this place because I want it to remain a hidden gem.  
This is a great find just outside of the hustle and bustle of downtown...
The Mercy Lounge hosts great bands and events, including local burlesque acts, and sports pin-up gals on the bar. 
Amazingly friendly staff. Building is just bursting with personality. 
Simply lovely. Every dish had an unusual but delicious mix of flavors and textures.

[Yelp reviews for touristy]
The signature iced tea was fabulous! There was a long wait but worth every minute of it! 
Reading Terminal Market has everything that you need everything is fresh good or good for you
The highlight of the cruse was the tour guide/narrator Charles and his eloquent oration. 
We were so excited to visit here from Las Vegas. What a Christmas experience for us!!!
It is very bustling so if you're one who doesn't like waiting in lines, you'll have to suck it up! 
If touristy is what you want, they serve great cheesesteaks!
Genos vibe was very touristy (we were tourists as well haha) but crowded. 
\end{lstlisting}

\clearpage

\section{Additional Computational Experiment Results}
\label{app:comp_results}
This appendix includes additional computational evaluation results for \textbf{E1} and \textbf{E2}, as well as an additional experiment \textbf{E3} for multi-preference steering. 

\renewcommand\thefigure{\thesection.\arabic{figure}} 
\setcounter{figure}{0}  

\renewcommand\thetable{\thesection.\arabic{table}} 
\setcounter{table}{0}  

\renewcommand\thelstlisting{\thesection.\arabic{lstlisting}} 
\setcounter{lstlisting}{0}  

\subsection{Additional Results for E1: Perplexity-Normalized Preference Effect}
\label{app:e1}
\begin{figure*}[h]
\vskip 0.1in
\begin{center}
\centerline{\includegraphics[width=\linewidth]{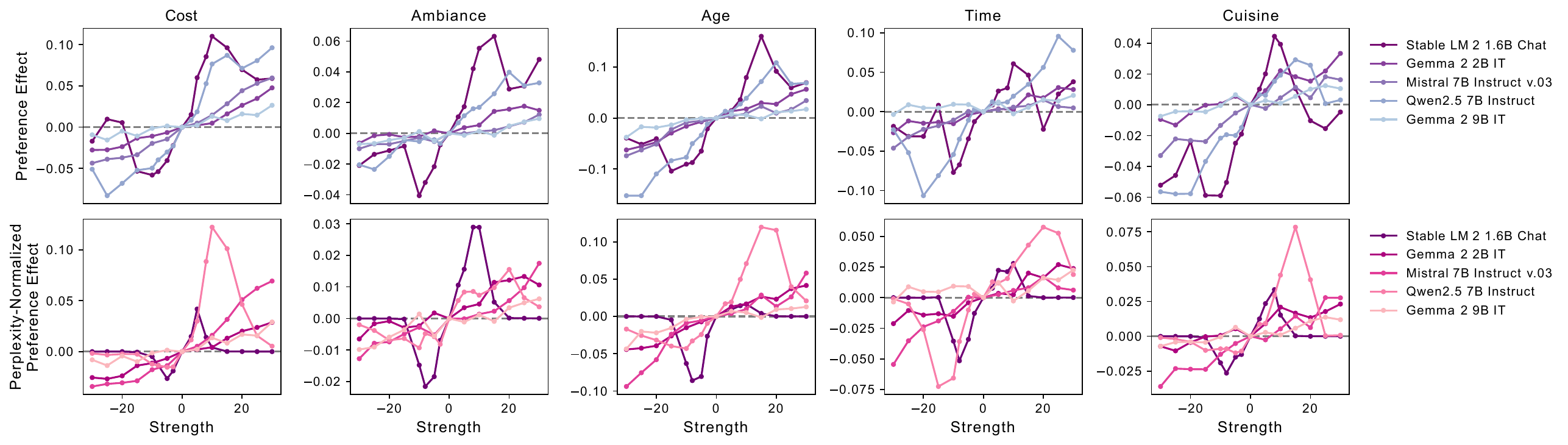}}
\caption{Effect of preferences (top) and perplexity-normalized effect (bottom) for all preferences and models for \textbf{(E1)}. }
\label{fig:pnes}
\end{center}
\end{figure*}

As described in the main text, we evaluate how steering strength modulates the expressed preferences of the model, including examining how model degeneracy impacts outputs with higher amounts of steering. To evaluate this, we compute the perplexity-normalized effect (PNE), which incorporates a measure of how the model's perplexity, a proxy for degeneration, changes with steering compared to a non-steered baseline. This equation is given by \citeauthor{DBLP:conf/icml/RutteABH24} as: $(\textit{Effect}_{steer=d} - \textit{Effect}_{steer=0}) / (\textit{PPL}_{steer=d} / \textit{PPL}_{steer=0})$, where PPL is perplexity and $d$ is the steering strength. If perplexity increases disproportionately more than the preference effect, then the PNE will be low, reflecting reduced quality. 
Figure \ref{fig:pnes} shows the standard (\textit{top}) and perplexity-normalized (\textit{bottom}) preference effects for all models and preferences. The values are standardized to center around $d=0$ as the neutral point with \textit{effect=0}. The PNE shows the appropximate stable ranges of steering for each model, which was mapped to the functional steering range in Table \ref{tab:params}.

\subsection{Additional Results for E2: Interactions Between Steering and Prompting}
\label{app:e2}
\begin{figure}[h]
\centering
\centerline{\includegraphics[width=\linewidth]{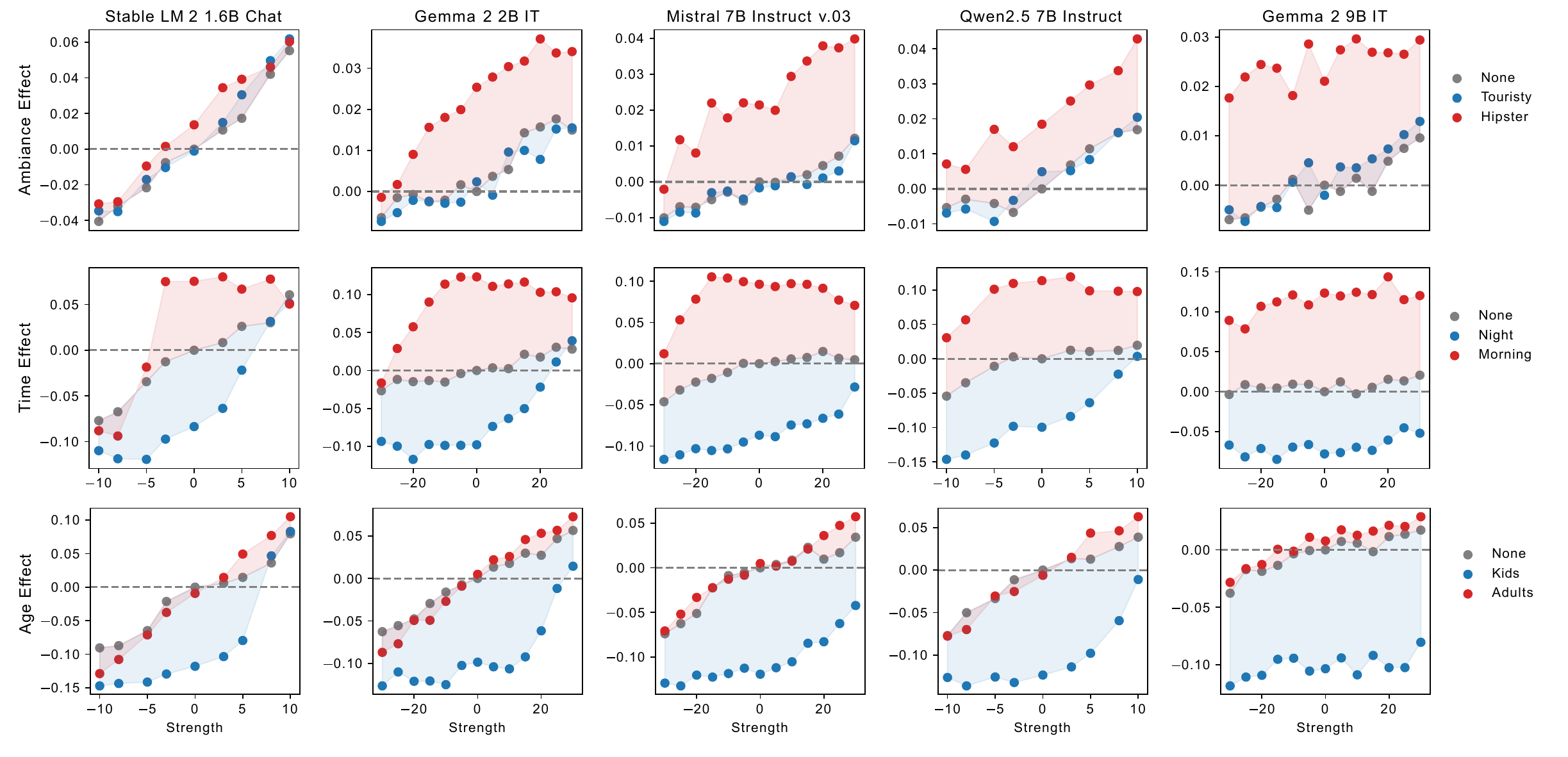}}
\caption{Effect of prompting on steering for preferences \textbf{ambiance} (top row), \textbf{time} (middle row), and \textbf{age} (bottom row) in \textbf{E2}. }
\label{fig:other_e2}
\end{figure}

The results for the effect of prompting in addition to steering for other preference dimensions in \textbf{E2} are in Figure \ref{fig:other_e2}. In particular, \textbf{time} is very sensitive to prompting, as evidenced by the large positive and negative offsets due to prompting towards \textit{morning} and \textit{night}. For the other dimensions, we observe that the default behaviour of the model for \textbf{ambiance} close aligns with \textit{touristy}, and \textbf{age} with \textit{adult}. This demonstrates what the default representations of the models are biased towards. Qualitative examples of how prompting impacts a \textbf{cost}-steered model are provided in Table \ref{tab:prompting}. For example, when the \textit{budget} user asks for more \textit{luxurious} options, then model outputs affordable suggestions that are reasonable for the user. 

\begin{table*}[h]
\caption{Samples of steered outputs with preference-based prompting for \textbf{cost} (E2).}
\label{tab:prompting}
\vskip -0.1in
\small
\begin{tabular}{p{0.14\linewidth}|p{0.25\linewidth}|p{0.25\linewidth}|p{0.25\linewidth}}
\toprule 
  \textbf{Cost} steering & 
  No prompting &
  Prompt towards \textit{luxury} &
  Prompt towards \textit{budget} \\ \midrule
  {\begin{minipage}{\linewidth} 
\vspace{0pt} ``Average User" (d=0)\end{minipage}} &
  {\begin{minipage}{\linewidth} \tiny  Here's a list of essential items you might want to consider for your camping trip. 1) Tent: Depending on the size of your family, you'll need a tent that can comfortably accommodate everyone...\end{minipage}}&
  {\begin{minipage}{\linewidth} \tiny  Here's a comprehensive list of camping supplies for your family camping holiday. 1) Tent: A spacious and high-quality tent is essential. Consider a tent with multiple rooms to provide privacy and separate areas for each family member...\end{minipage}} &
  {\begin{minipage}{\linewidth}   \tiny  Here's a list of essential camping supplies to help you get started while keeping it budget-friendly. 1) Tent: A 4-person dome tent is a popular and affordable choice...\end{minipage}}\\ \midrule
 {\begin{minipage}{\linewidth} ``Wealthy User" (d=20) \end{minipage}}&
  {\begin{minipage}{\linewidth} \tiny To ensure your comfort and convenience, I have prepared a comprehensive list of essentials. 1) Tent: A four-season, safari-style tent with a large canopy...\end{minipage}} &
  {\begin{minipage}{\linewidth} \tiny To ensure that your stay is nothing less than perfection, I have prepared a meticulously curated list of the most exquisite and indulgent camping supplies...\end{minipage}} &
  {\begin{minipage}{\linewidth} \tiny To ensure a comfortable and enjoyable adventure, I have compiled a list of essential items that will cater to your needs without exceeding your budget. 1) Tent: A spacious 4-person dome tent with a waterproof flysheet and a mesh canopy for stargazing...\end{minipage}} \\ \midrule
  {\begin{minipage}{\linewidth} ``Budget User" (d=-20) \end{minipage}}&
  {\begin{minipage}{\linewidth} \tiny Tent: Depending on your camping style, you may need a tent. There are various types of tents available, such as dome tents, backpacking tents, and family tents...\end{minipage}} &
  {\begin{minipage}{\linewidth} \tiny Here are some essential items to consider when planning a camping trip: 1) Tent: There are various types of tents available, from simple pop-up tents to more complex backpacking tents...\end{minipage}} &
  {\begin{minipage}{\linewidth} \tiny Here are some budget-friendly supplies for a camping holiday: 1) Tent: Consider buying used tents or borrowing from friends...\end{minipage}}\\
  \bottomrule
\end{tabular}
\end{table*}

\clearpage

\subsection{Additional Results for E3: Multi-Preference Steering}
\label{app:e3}

\begin{figure}[h]
\centering
\centerline{\includegraphics[width=\linewidth]{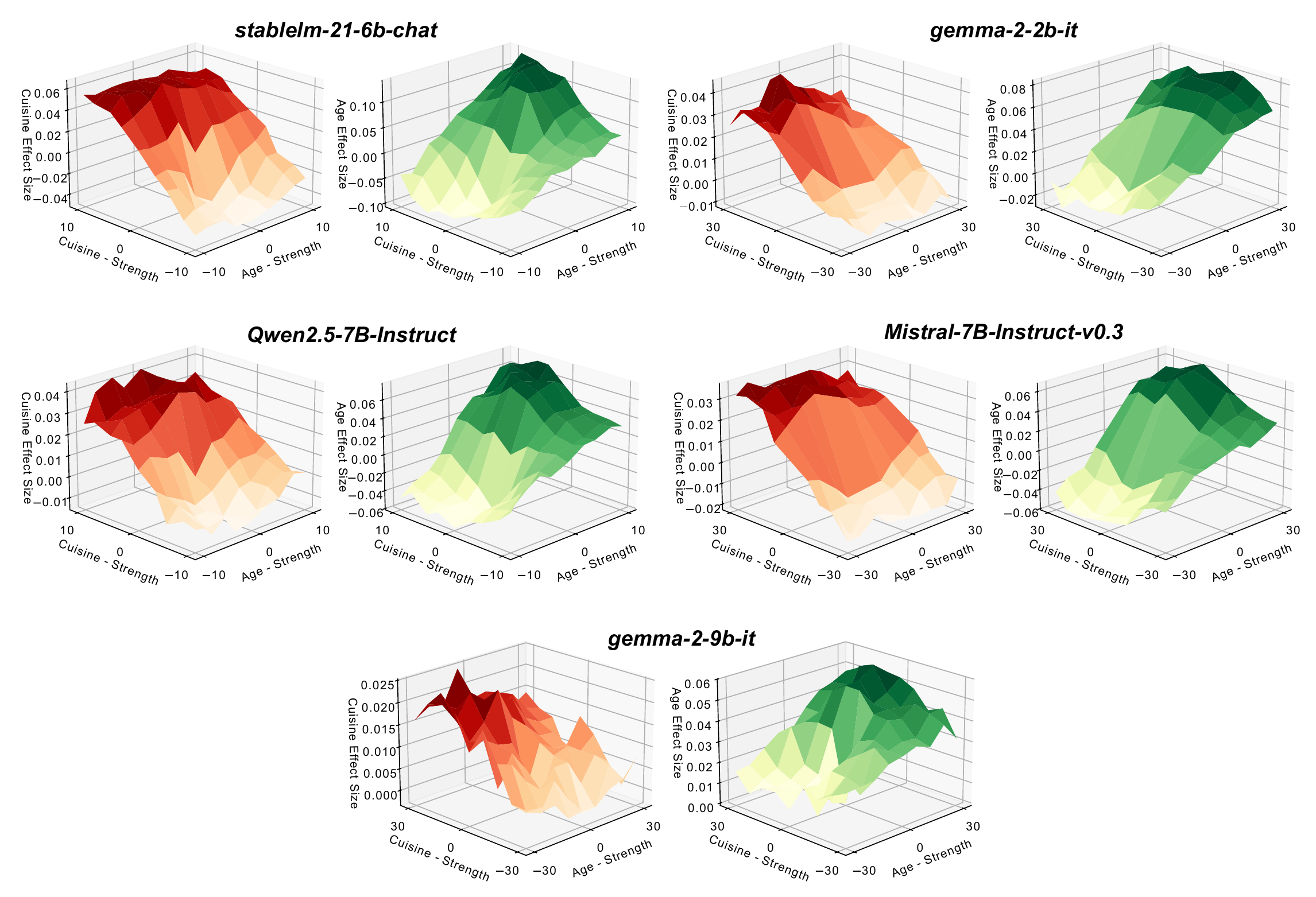}}
\caption{Additional multi-preference steering surface plots for \textbf{E3}. }
\label{fig:app_e3_figs}
\end{figure}

In this additional experiment on the preliminary implementation of multi-preference steering, we focus on two dimensions instead of all five at once. In future context, this should be extended to incorporate more preferences, ideally ones that are are orthogonal to each other. 
Since our preferences are not inherently orthogonal, we select two dimensions with the least collinearity, which are \textbf{cuisine} and \textbf{age}.
We vary the steering strength for each dimension uniformly and measure the effect of each preference separately. The steering vector is applied as a weighed combination of the individual preference vectors.
With compounded steering, the main concern is that the effects will deteriorate. Figure \ref{fig:app_e3_figs} shows 3D plots of the variation of the preference effect for \textbf{cuisine} and \textbf{age}. While the range of effects is slightly degraded at large values of complementary steering, the surfaces follow the expected trend. Qualitative samples of multi-steered texts are shown in Table \ref{tab:multisteer}, which highlights phrases that exemplify each of the target traits. The results of \textbf{E3} suggest that \textbf{\textit{multiple preferences can be steered additively}}, which is promising for scaling up steering as a personalization \edit{strategy}.


\begin{table*}[h]
\caption{Samples from multi-preference steering with preferences \textbf{age} and \textbf{cuisine} (E3).}
\label{tab:multisteer}
\vskip -0.1in
\small
\begin{tabular}{p{0.13\linewidth}|p{0.4\linewidth}|p{0.4\linewidth}}
\toprule 
 & 
  {\begin{minipage}{\linewidth} 
\vspace{0pt}Mistral 7B ($d_{cuisine}=30$, steer $\rightarrow$ \hlc[pink]{American})\end{minipage}} &
  {\begin{minipage}{\linewidth} 
\vspace{0pt}Mistral 7B ($d_{cuisine}=-30$, steer $\rightarrow$ \hlc[cyan]{Asian})\end{minipage}} \\ \midrule
  {\begin{minipage}{\linewidth} 
\vspace{0pt}Mistral 7B ($d_{age}=30$, steer $\rightarrow$ \hlc[lime]{Adults})\end{minipage}} &
  {\begin{minipage}{\linewidth} \tiny  San Francisco is known for its vibrant dining scene, offering a \hlc[lime]{wide range of culinary experiences}. Here are some of the top-rated restaurants in the city. 1) The Slanted Door - Overlooking the Bay, this restaurant offers breathtaking views alongside a menu of contemporary \hlc[pink]{American dishes}...\end{minipage}}&
  {\begin{minipage}{\linewidth} \tiny  San Francisco, being a vibrant city with a \hlc[lime]{rich culinary scene}, offers a wide array of exceptional restaurants that cater to various gastronomic preferences. 1) Izakaya Roku: Known for its authentic and vibrant \hlc[cyan]{izakaya} atmosphere, this restaurant offers a comprehensive selection of \hlc[cyan]{Japanese small dishes and sake...}\end{minipage}} \\ \midrule
  {\begin{minipage}{\linewidth} 
\vspace{0pt}Mistral 7B ($d_{age}=-30$, steer $\rightarrow$ \hlc[yellow]{Kids}) \end{minipage}} &
  {\begin{minipage}{\linewidth} \tiny  San Francisco offers a variety of \hlc[yellow]{family-friendly} restaurants that are popular among tourists and locals. Here are some of the best restaurants in the San Francisco Bay Area. 1) \hlc[pink]{The Original Pancake House}: A roadside attraction featuring \hlc[yellow]{giant dinosaur statues} and a restaurant serving \hlc[pink]{breakfast foods}...\end{minipage}}&
  {\begin{minipage}{\linewidth} \tiny Here are some popular restaurants that you can consider visiting. 1) Golden Gate \hlc[cyan]{Fortune Cookie Factory}: This is a \hlc[yellow]{fun and educational} experience where you can make your own \hlc[cyan]{fortune cookies}. It's a great way to \hlc[yellow]{introduce kids} to different shapes and designs...\end{minipage}} \\ \bottomrule

\end{tabular}
\end{table*}

\clearpage

\section{Additional User Study Details}
\label{app:user_study}

\renewcommand\thefigure{\thesection.\arabic{figure}} 
\setcounter{figure}{0}  

\renewcommand\thetable{\thesection.\arabic{table}} 
\setcounter{table}{0}  

\renewcommand\thelstlisting{\thesection.\arabic{lstlisting}} 
\setcounter{lstlisting}{0}  

This appendix provides the details of implementing the user study as well as additional results not covered by the main text. 


\subsection{Interfaces Screenshots}
\label{app:screenshots}
\begin{figure*}[h!]
    \centering
    \begin{subfigure}[b]{0.32\textwidth}
        \includegraphics[width=\linewidth]{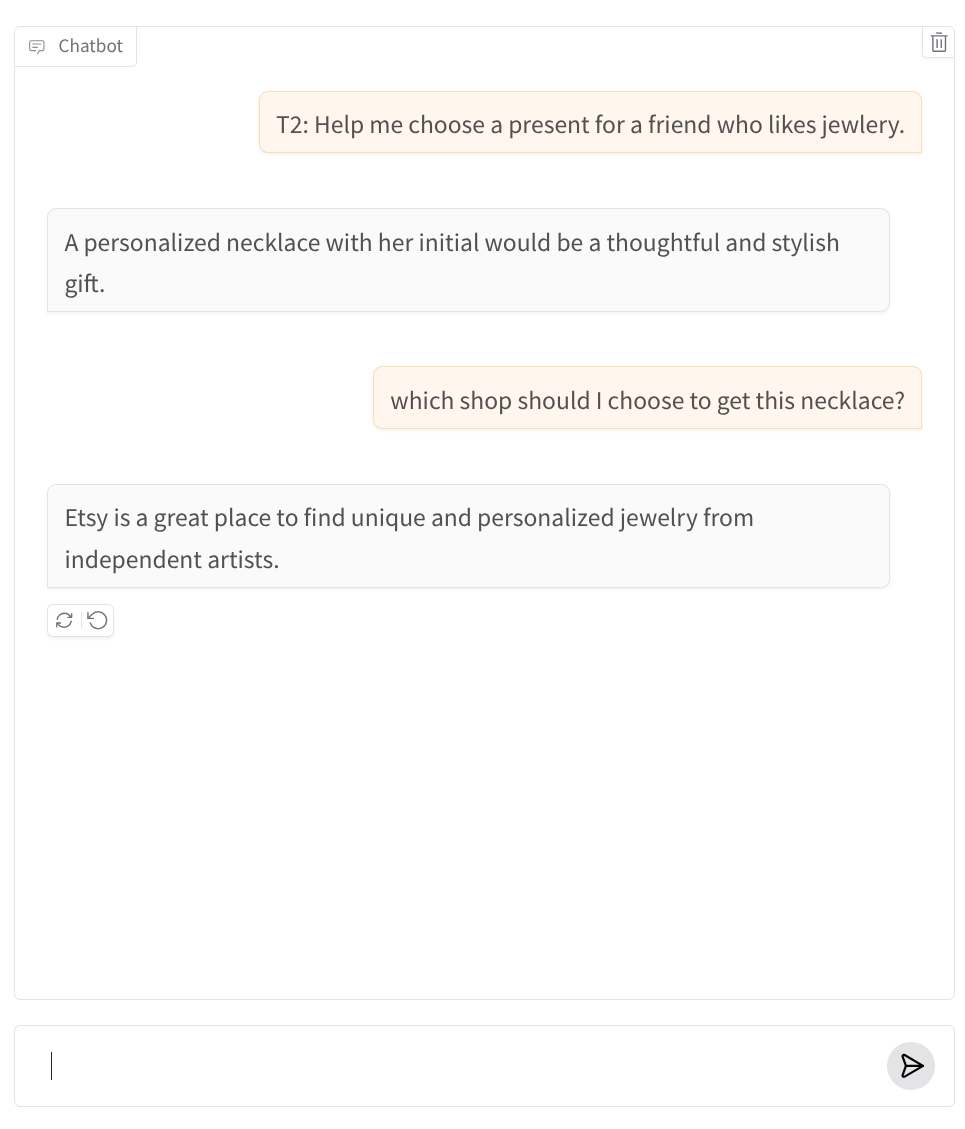}
        \caption{\prompt{} baseline interface.}
        \label{fig:prompt}
    \end{subfigure}
    \hfill
    \begin{subfigure}[b]{0.32\textwidth}
        \includegraphics[width=\linewidth]{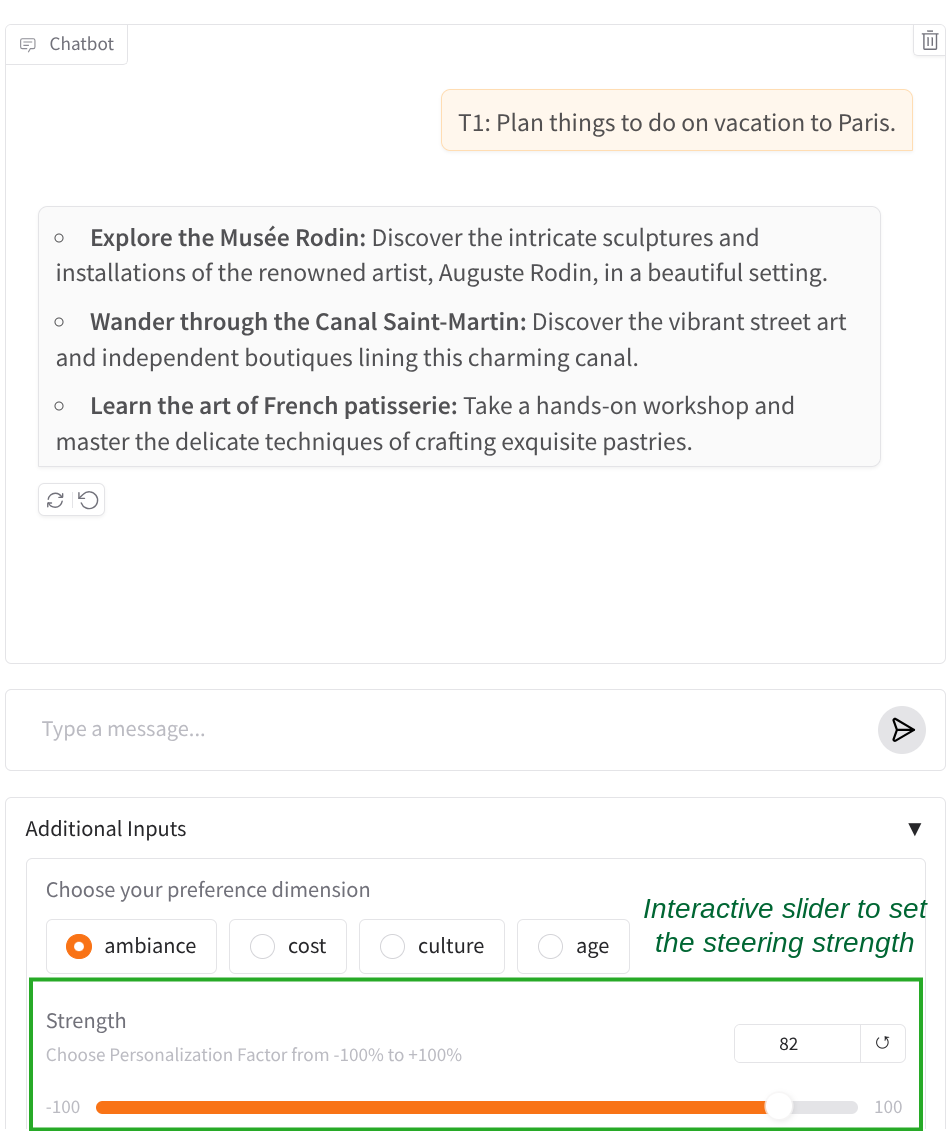}
        \caption{\select{} interface with preference slider.}
        \label{fig:select}
    \end{subfigure}
    \hfill
    \begin{subfigure}[b]{0.32\textwidth}
        \includegraphics[width=\linewidth]{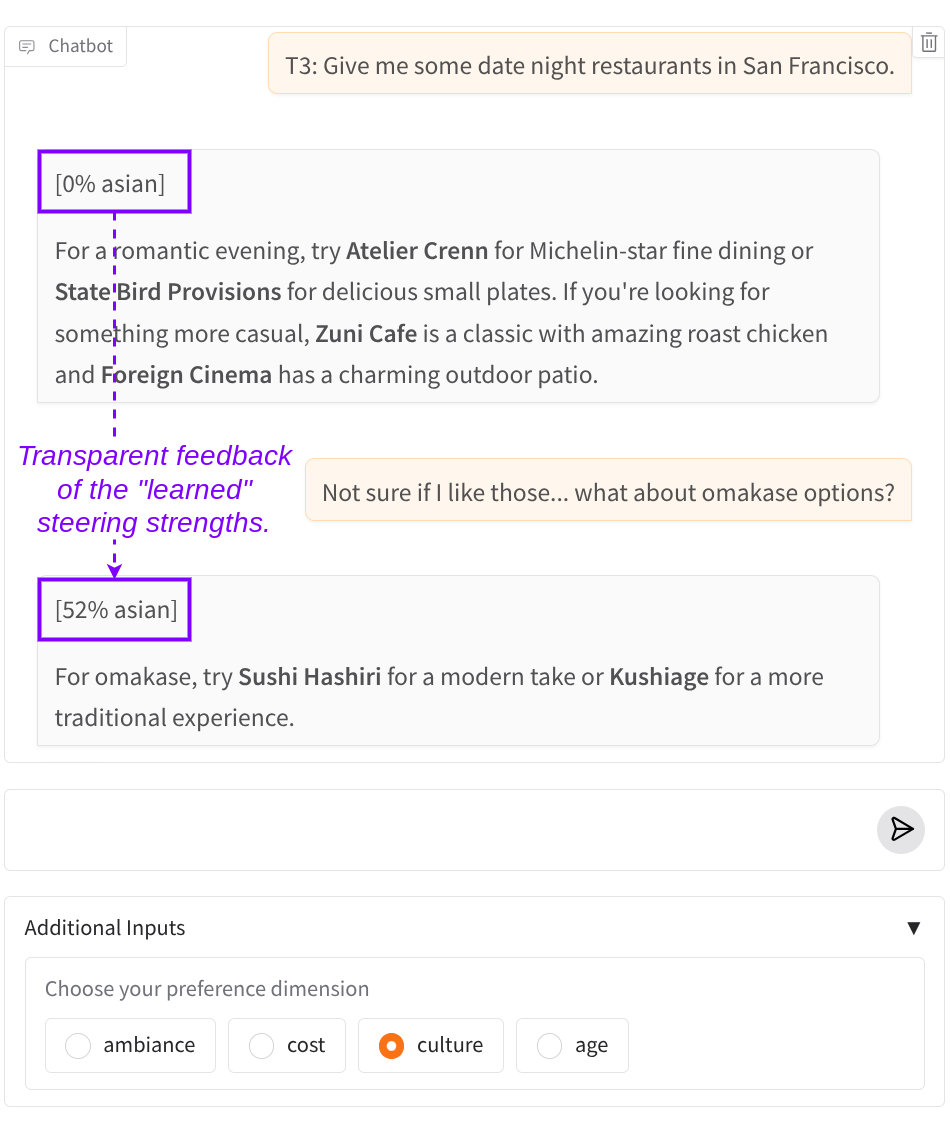}
        \caption{\learn{} interface with inferred preference.}
        \label{fig:learn}
    \end{subfigure}
    \\ \vspace{0.3in}
    \begin{subfigure}[b]{0.41\textwidth}
        \includegraphics[width=\linewidth]{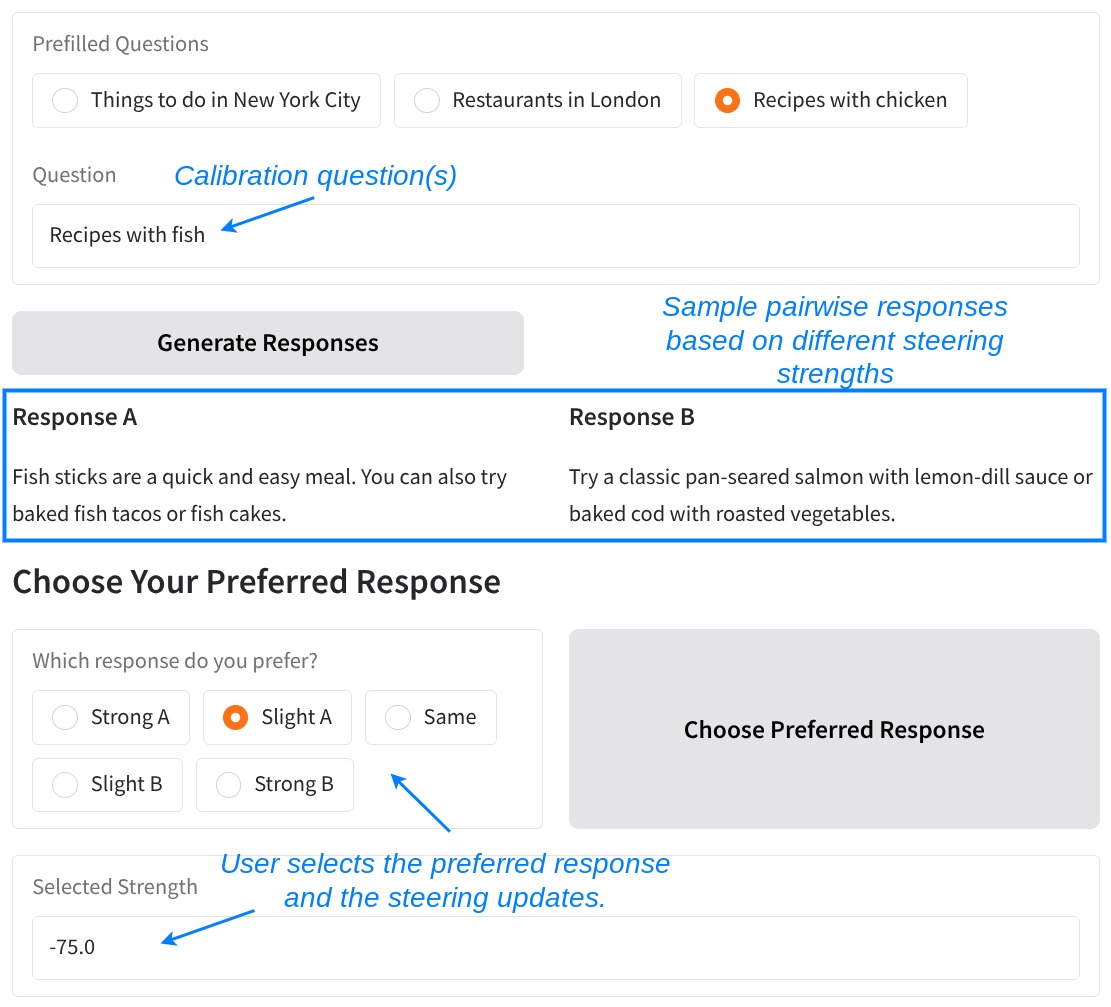}
        \caption{Calibration step of \calibrate{}.}
        \label{fig:calibrate1}
    \end{subfigure}
    \hspace{0.3in}
    \begin{subfigure}[b]{0.33\textwidth}
        \includegraphics[width=\linewidth]{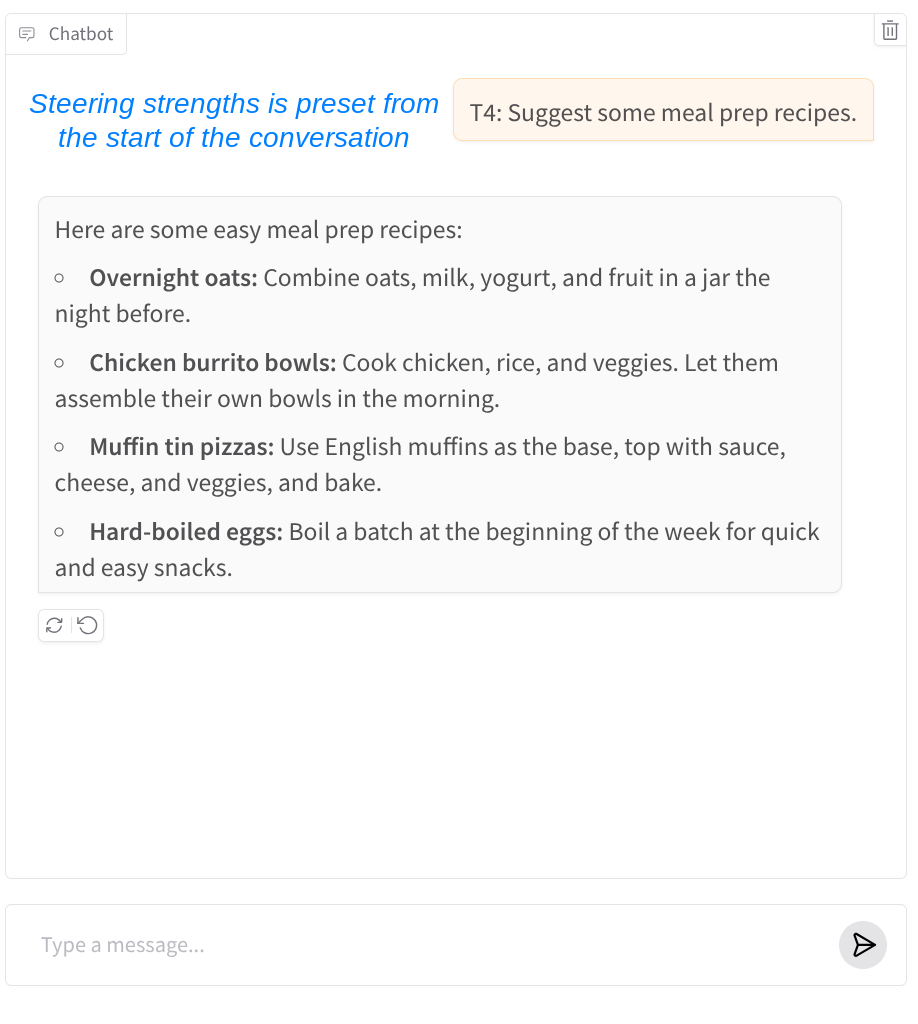}
        \caption{\calibrate{} chat with preset preference.}
        \label{fig:calibrate2}
    \end{subfigure}
    \caption{Screen captures of all four chatbot interfaces that participants used.}
    \label{fig:interfaces_screenshots}
\end{figure*}

As a correspondence to the interface diagrams in Figure \ref{fig:interfaces}, we also provide screenshots of what the participants actually saw in Figure \ref{fig:interfaces_screenshots}. The interfaces were implemented using Gradio \cite{abid2019gradio}. 

\clearpage
\subsection{Participants' Ground Truth Preferences}
\label{app:gt_preferences}

\begin{figure}[h]
\vskip -0.1in
\begin{center}
\centerline{\includegraphics[width=0.35\linewidth]{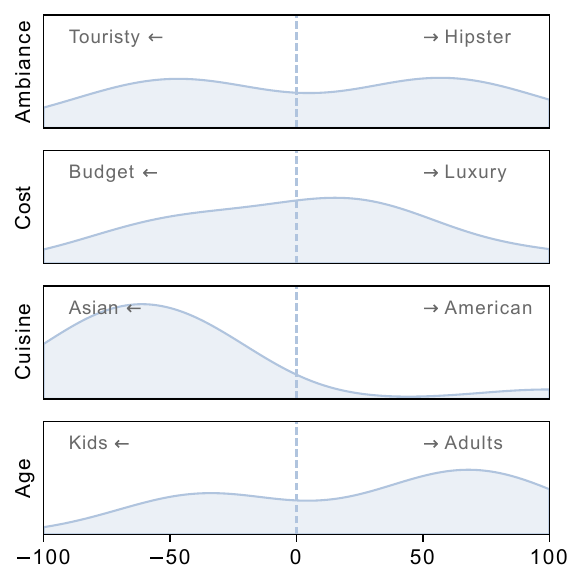}}
\vskip -0.1in
\caption{Ground truth preference distributions.}
\label{fig:pref_dist}
\end{center}
\vskip -0.1in
\end{figure}

Figure \ref{fig:pref_dist} shows the distribution of true preferences across each of the four tasks/preferences pairings as they were self-reported by the user study participants. The preferences of \textbf{ambiance} and \textbf{cost} are well-distributed across both positive and negative traits, while \textbf{cuisine} leaned towards \textit{Asian} and \textbf{age} leaned towards \textit{adults}.

\subsection{\learn{} Implementation and Experimental Outcomes}
\label{app:learned_strengths}

We provide more detail on how the \learn{} algorithm is implemented. As presented in the main text, the steering strength is updated based on the content and sentiment of the user's message: \setlength{\abovedisplayskip}{3pt}
\setlength{\belowdisplayskip}{3pt}
\begin{equation}
\label{eq:update}
    d^{u*}_{t+1} \leftarrow d^{u*}_{t} + 
    p( \mathrm{dissatisfaction}(x_t) \cdot \mathrm{direction}(x_t) )
\end{equation}

Here, $x_t$ is the user's message at round \textit{t} of the conversation, $dissatisfaction(\cdot)$ measures the granular sentiment of dissatisfaction in the user’s feedback, $direction(\cdot)$ is a binary classification that determines whether the user wants the model’s outputs to shift positively or negatively, and $p(\cdot)$ is a simple linear transformation. 
While there are many possible choices for carrying out these computations, we demonstrate the aptitude of the high-level methodology using straightforward choices:
\begin{itemize}[topsep=4pt, noitemsep]
    \item $dissatisfaction(\cdot)$ is taken as the weighted sum of the prediction probabilities of negative and neutral sentiment detected with the pretrained classifier TweetEval \cite{barbieri2020tweeteval}\footnote{We use a weighted computation of $0.75 * p_{Negative} + 0.25 * p_{Neutral}$ because it offers greater discrimination than just the negative probability alone.}.
    \item $direction(\cdot)$ is determined based on the greater cosine similarity to Sentence-BERT \cite{reimers-2019-sentence-bert} embeddings of reference phrases (e.g., \textit{``I want more luxury"} and \textit{``I want lower cost"} for the preference dimension of \textbf{cost}).
    \item $p(\cdot)$ is a remapping of the estimated preference value, for example, to the functional steering range of the model.
\end{itemize}

\begin{figure}[h!]
\vskip -0.1in
\begin{center}
\centerline{\includegraphics[width=0.45\linewidth]{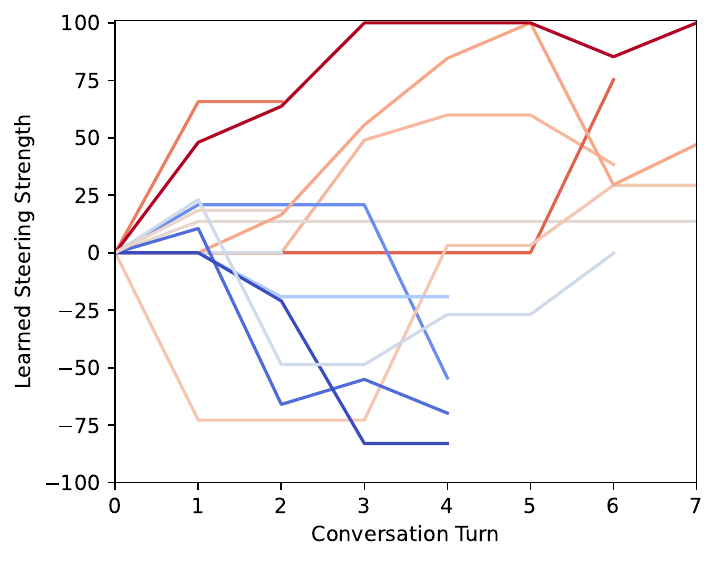}}
\vskip -0.1in
\caption{Progression of strengths learned by \learn{} across all conversations. }
\label{fig:learned_strenghts}
\end{center}
\vskip -0.1in
\end{figure}

In the user study itself, Figure \ref{fig:learned_strenghts} shows the progression of the learned steering strengths across all 14 participant conversations, with red indicating more positive strengths and blue indicating more negative. 

\subsection{Error in Steering Strengths}
\label{app:error}

\begin{figure}[h]
\vskip -0.1in
\begin{center}
\centerline{\includegraphics[width=0.6\linewidth]{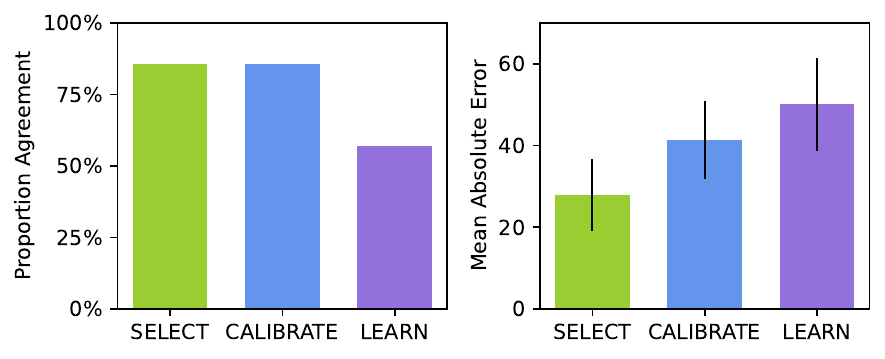}}
\vskip -0.1in
\caption{\edit{Proportion of agreement between the applied steering direction with the} participants' true preference (left) and mean absolute error of the steering strengths (right).}
\label{fig:steering_stats}
\end{center}
\vskip -0.1in
\end{figure}

For \textbf{U1}, we also examine if the applied steering strengths match the ground truth preferences of the participants. For \select{}, we take the steering strength set by the participant in the first step; for \calibrate{}, we use the post-calibration steering strength; and for \learn{}, we use the learned strength in the last turn of the conversation. Figure \ref{fig:steering_stats} compares the chatbots across two measures --- do the steering strengths directionally agree with ground truth preferences (e.g. is the steering value \textit{positive} if the user indicates they prefer \textit{luxury} for \textbf{cost}?); and what is the absolute error between the steering value and ground truth for the given task and preference? 

\edit{Averaged across all participants}, \select{} had the lowest mean absolute error ($M=27.8, SD=31.9$) and highest proportion of agreement at 86\% --- although it is surprisingly not 100\%, as participants chose it themselves. \calibrate{} matched the same agreement rate, but had higher error ($M=41.4, SD=34.6$). \learn{} was less robust and achieved slightly above random rate of agreement at 57\%, with the highest error ($M=50.1, SD=40.9$)\edit{, again reflecting the potential weakness in the learning algorithm. In summary, we observe varying levels of accuracy in the steerable chatbots’ alignment with ground-truth preferences, but we contend that future advances in calibration and learning algorithms can substantially improve this performance.} 

\end{document}